\documentclass[12pt, draftclsnofoot, onecolumn]{IEEEtran}

\usepackage[cmex10]{amsmath}
\usepackage{amssymb}

\usepackage{url}

\usepackage{cases}
\usepackage{multirow}
\usepackage{empheq}

\usepackage{array}

\usepackage[noadjust]{cite}

\usepackage{verbatim}

\ifCLASSINFOpdf
  \usepackage[pdftex]{graphicx}
\else
  \usepackage[dvips]{graphicx}
\fi

\usepackage{color}

\newtheorem{theorem}{Theorem}
\newtheorem{lemma}[theorem]{Lemma}

\newtheorem{corollary}[theorem]{Corollary}

\newtheorem{remark}[theorem]{Remark}

\DeclareMathOperator{\rank}{rank}

\begin{document}
%
\title{On the Duality of Erasures and Defects}
%
%
%

\author{Yongjune~Kim,~\IEEEmembership{Student Member,~IEEE} and~B.~V.~K.~Vijaya~Kumar,~\IEEEmembership{Fellow,~IEEE}
\thanks{Parts of the material in this paper were presented in part at the IEEE International Conference on Communications, Budapest, Hungary, June 2013 and the IEEE International Symposium on Information Theory, Istanbul, Turkey, July 2013.

Y. Kim and B. V. K. Vijaya Kumar are with the Department
of Electrical and Computer Engineering, Carnegie Mellon University, Pittsburgh,
PA, 15213, USA (e-mail: yongjunekim@cmu.edu, kumar@ece.cmu.edu).}
}

\maketitle

\begin{abstract}
In this paper, the duality of erasures and defects will be investigated by comparing the binary erasure channel (BEC) and the binary defect channel (BDC). The duality holds for channel capacities, capacity achieving schemes, minimum distances, and upper bounds on the probability of failure to retrieve the original message. Also, the binary defect and erasure channel (BDEC) will be introduced by combining the properties of the BEC and the BDC. It will be shown that the capacity of the BDEC can be achieved by the coding scheme that combines the encoding for the defects and the decoding for the erasures. This coding scheme for the BDEC has two separate redundancy parts for correcting erasures and masking defects. Thus, we will investigate the problem of redundancy allocation between these two parts.

\end{abstract}


%
\IEEEpeerreviewmaketitle

\section{Introduction}
%
%
%
%


The binary erasure channel (BEC) is a very well known channel of \emph{communication} which was introduced by Elias in 1955 \cite{Elias1955}. Due to its simplicity, it has been a starting point to design new coding schemes and analyze the properties of codes \cite{Arikan2009, Di2002}. In addition, coding schemes for BEC are still being actively researched since BEC is a very good model of the Internet \cite{Luby2001, Luby2002, Shokrollahi2006}.

\begin{figure}[!t]
   \centering
   \includegraphics[width=0.30\textwidth]{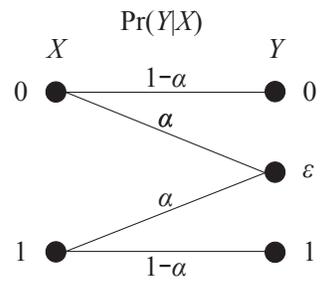}
   \caption{Binary erasure channel (BEC).}
   \label{fig:BEC}
\end{figure}

In BEC, as shown in Fig.~\ref{fig:BEC}, the channel input $X \in \{0, 1\}$ is binary and the channel output $Y=\{0, 1, \varepsilon\}$ is ternary. It is assumed that the decoder knows the locations of erased bits denoted by $\varepsilon$. The capacity of the BEC with erasure probability $\alpha$ is given by \cite{Elias1955, Cover2006}
\begin{equation}\label{eq:BEC_capacity}
C_{\mathrm{BEC}} = 1 - \alpha.
\end{equation}

Elias \cite{Elias1955} showed that random codes of rates arbitrarily close to $C_{\textrm{BEC}}$ can be decoded on the BEC with an exponentially small error probability using maximum likelihood (ML) decoding. In the case of BEC, ML decoding of linear codes is equivalent to solving linear equations \cite{Elias1955, Shokrollahi2006}.

The binary defect channel (BDC) also has a long history. The BDC was introduced to model computer memory for \emph{storage} by Kuznetsov and Tsybakov in 1974 \cite{Kuznetsov1974}. At that time, erasable and programmable read only memories (EPROM) and random access memories (RAM) were modeled by the BDC \cite{Kuznetsov1974}. Recently, BDC has received renewed attention for nonvolatile memories such as flash memories and phase change memories (PCM) \cite{Hwang2011a, Jagmohan2010a, Lastras-Montano2010}. In addition, BDC is theoretically important since write once memories (WOM), write unidirectional memories (WUM), and some other constrained memories can be considered as special cases of the BDC \cite{Kuzntsov1994}.

\begin{figure}[!t]
   \centering
   \includegraphics[width=0.60\textwidth]{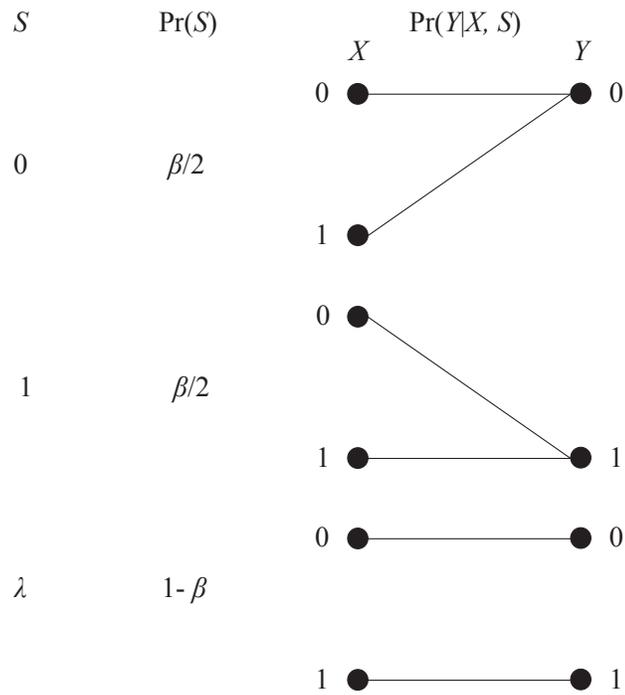}
   \caption{Binary defect channel (BDC).}
   \label{fig:BDC}
\end{figure}

As shown in Fig.~\ref{fig:BDC}, BDC has a ternary channel state $S \in \{0, 1, \lambda\}$ whereas the channel input $X$ and the channel output $Y$ are binary. The state $S=0$ corresponds to a stuck-at 0 defect that always outputs a 0 independent of its input value, the state $S=1$ corresponds to a stuck-at 1 defect that always outputs a 1, and the state $S = \lambda$ corresponds to a normal cell that outputs the same value as its input. The probabilities of these states are $\beta / 2$, $\beta / 2$ (assuming a symmetric defect probability), and $1 - \beta$, respectively  \cite{ElGamal2011, Heegard1983:capacity, Heegard1983}.

It is known that the capacity is $1 - \beta$ when both the encoder and the decoder know the defect information. If the decoder is aware of the defect locations, then the defects can be regarded as erasures so that the capacity is $1 - \beta$ \cite{ElGamal2011, Heegard1983:capacity}. On the other hand, Kuznetsov and Tsybakov assumed that the encoder knows the defect information such as locations and stuck-at values of defects and the decoder does not have any information of defects \cite{Kuznetsov1974}. It was shown that the capacity is also $1 - \beta$ even if only the encoder knows the defect information \cite{ElGamal2011, Heegard1983:capacity}. Thus, the capacity of the BDC is given by
\begin{equation}\label{eq:BDC_capacity}
C_{\mathrm{BDC}} = 1 - \beta.
\end{equation}


The capacity of the BDC can be achieved by \emph{random binning} when only the encoder knows the defect information \cite{Heegard1983:capacity, ElGamal2011}. The practical coding scheme is the \emph{additive encoding} which masks defects by adding a carefully selected binary vector \cite{Kuznetsov1974, Tsybakov1975b}. Masking defects is to make a codeword whose values at the locations of defects match the stuck-at values of the defects at those locations.

Heegard proved that the additive encoding (formulated as an optimization problem) and ML decoding can achieve the capacity of a channel that has both defects and random errors \cite{Heegard1983}. Note that the additive encoding masks defects and the ML decoding corrects random errors.

In \cite{Kim2013}, an upper bound on the probability of masking failure was derived when the additive encoding is accomplished by solving the linear equations instead of solving the optimization problem. The derived upper bound is based on the weight distribution of the underlying codes. Based on the upper bound of \cite{Kim2013}, we will show that the additive encoding can achieve $C_{\mathrm{BDC}}$ by using random linear codes and solving a system of linear equations. In addition, numerical results show that structured linear codes such as Bose, Chaudhuri, and Hocquenghem (BCH) codes are good choices since their performance is not far from $C_{\mathrm{BDC}}$.

In Section~\ref{sec:duality}, the BEC and the BDC will be discussed separately. Their channel properties, capacities, capacity achieving coding schemes and upper bounds on the probability of failure will be discussed comprehensively. Afterwards, we will investigate the duality of the BEC and the BDC. Basically, an erasure $\varepsilon$ is \emph{neither} 0 nor 1. In contrast, a stuck-at value (i.e., defect value) is \emph{either} 0 or 1. Also, the \emph{decoder} corrects erasures in the BEC and the \emph{encoder} masks defects in the BDC. Both channels have similar capacities as shown in \eqref{eq:BEC_capacity} and \eqref{eq:BDC_capacity}. In addition, both capacities can be achieved by solving the linear equations.

However, as we will show later in this paper, the BEC and the BDC have some important differences. The linear equations for the BEC can be described by an \emph{overdetermined} system and there is only one solution that corrects all erasures. Meanwhile, the linear equations for the BDC correspond to an \emph{underdetermined} system which allows several solutions that mask defects. In addition, the solution of linear equations for the BEC is the estimate of message or the estimate of erased bits, whereas the solution for the BDC is the parity or the the codeword.

In addition, the minimum distance and the weight distribution of the coding scheme for the BEC are controlled by the \emph{parity check matrix}, whereas the minimum distance and the weight distribution of the coding scheme for the BDC come from the \emph{generator matrix}. Because of these duality properties, the upper bound on the probability of decoding failure for the BEC and the upper bound on the probability of masking failure for the BDC have interesting similarities and differences. Considering that the BEC is a channel model for digital \emph{communication} and the BDC is a channel model for digital \emph{storage}, the duality is meaningful.

\begin{figure}[!t]
   \centering
   \includegraphics[width=0.60\textwidth]{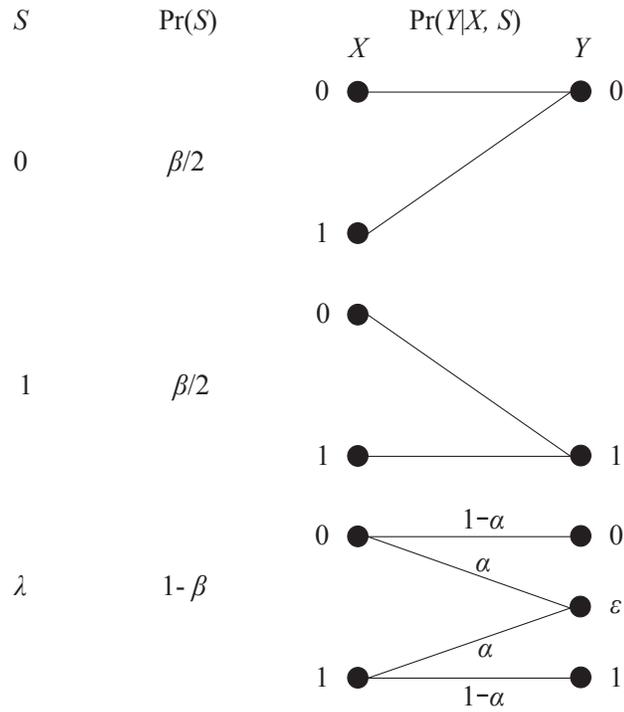}
   \caption{Binary defect and erasure channel (BDEC).}
   \label{fig:BDEC}
\end{figure}

In Section~\ref{sec:BDEC}, the binary defect and erasure channel (BDEC) will be introduced. As shown in Fig.~\ref{fig:BDEC}, the BDEC has both erasures (with erasure probability $\alpha$ for a normal cell) and defects (with defect probability $\beta$). The capacity of the BDEC is given by
\begin{equation}\label{eq:BDEC_capacity}
C_{\mathrm{BDEC}} = \left(1 - \alpha \right)\left(1 - \beta \right).
\end{equation}

We will show that the capacity of the BDEC can be achieved by a coding scheme that combines the encoding of the BDC and the decoding of the BEC. This proposed coding scheme for BDEC has two separated redundancy parts: one for correcting erasures and the other for masking defects. In order to minimize the probability of failure of correcting erasures and masking defects, the redundancy allocation between these two redundancy parts should be optimized.

During our proof that the proposed coding scheme achieves the capacity $C_{\mathrm{BDEC}}$, lower bounds on these two redundancy components (achieving the capacity) can be obtained. However, these lower bounds may not be of much help in determining the redundancy allocation when the codeword length is finite.

Thus, we will investigate redundancy allocation for the BDEC. First, the optimal redundancy allocation is obtained by simulations. Then, we will derive the upper bound on the probability of failure for a finite codeword length and obtain the estimate of the optimal redundancy allocation by minimizing this upper bound instead of the probability of failure. Same methodology has been applied for the channel that has both defects and random errors in \cite{Kim2013:plbc} and it was shown that the estimated redundancy allocation matches the optimal one very well. From the numerical results, we will show that this method to minimize the upper bound works well for the BDEC as well as the channel of defects and random errors.

The rest of the paper is as follows. Section~\ref{sec:duality} discusses the duality between erasures and defects. In Section~\ref{sec:BDEC}, the BDEC will be introduced and the redundancy allocation for the BDEC will be investigated. After discussing the numerical results in Section~\ref{sec:numerical_results}, we conclude in Section~\ref{sec:conclusion}.

\section{Duality between Erasures and Defects} \label{sec:duality}

%
%
%

\subsection{Binary Erasure Channel}

For the BEC, the codeword most likely to have been transmitted is the one that agrees with all of received bits that have not been erased. If there is more than one such codeword, the decoding may lead to a failure. Thus, the following simple coding scheme was proposed in \cite{Elias1955}.

\emph{Encoding:} A message $\mathbf{m}\in \{0, 1\}^k$ is encoded to a corresponding codeword $\mathbf{c} \in \{0, 1\}^n$ by $\mathcal{C} = \left\{\mathbf{c} = G \mathbf{m} \mid \mathbf{m} \in \{0, 1\}^k \right\}$ where $\mathcal{C}$ is a set of codewords and the generator matrix $G$ is an $n \times k$ matrix over $\{0, 1\}$ such that $\rank(G)=k$. Note that the code rate $R = \frac{k}{n}$.

\emph{Decoding:} Let $g: \mathbf{y} \in \{0, 1, \varepsilon \}^n \rightarrow \mathcal{C}$ denote the decoding rule. If the channel output $\mathbf{y}$ is identical to one and only one codeword on the unerased bits, the decoding succeeds. If $\mathbf{y}$ matches completely with several codewords on the unerased bits, the decoder chooses one of them randomly. Note that there exists at least one codeword that matches with $\mathbf{y}$ on the unerased bits \cite{Elias1955}.

We will define a random variable $D$ as follows.
\begin{equation}
D =
\begin{cases}
0,  & \mathbf{c} \ne \widehat{\mathbf{c}}\text{ (decoding failure)}; \\
1,  & \mathbf{c} = \widehat{\mathbf{c}}\text{ (decoding success)}
\end{cases}
\end{equation}
where $\widehat{\mathbf{c}}$ is the estimated codeword produced by the decoding rule of $g$.

The minimum distance $d_{\mathrm{min}}$ of $\mathcal{C}$ is given by
\begin{align} \label{eq:BEC_dmin}
d_{\mathrm{min}} &= \underset{
\substack{
\mathbf{c} \ne \mathbf{0} \\
H^T \mathbf{c}= \mathbf{0}
}}
{\text{min }} \|\mathbf{c}\|
\end{align}
where the parity check matrix $H$ is an $n \times (n-k)$ matrix such that $H^{T}G = \mathbf{0}$ (superscript $T$ denotes transpose). Also, $\| \cdot \|$ represents the Hamming weight of a vector. Due to \eqref{eq:BEC_dmin}, any $d_{\mathrm{min}} - 1$ rows of $H$ are linearly independent. If $e < d_{\mathrm{min}}$, all $e$ erasures will be successfully corrected, which will be shown in Lemma~\ref{lemma:BEC_UB}. 

The decoding rule of $g$ can be described by the following linear equations \cite{Elias1955}.
\begin{equation}\label{eq:BEC_decoder_LE}
G^{\mathcal{V}} \widehat{\mathbf{m}} = \mathbf{y}^{\mathcal{V}}
\end{equation}
where $\widehat{\mathbf{m}}$ is the estimate of $\mathbf{m}$ and $\mathcal{V}=\left\{j_1,\cdots, j_v\right\}$ indicates the locations of $v$ unerased bits. We use the notation of $\mathbf{y}^{\mathcal{V}}=\left(y_{j_1}, \cdots, y_{j_v}\right)^T$ and $G^{\mathcal{V}}=\left[ \mathbf{g}_{j_1}^T, \cdots, \mathbf{g}_{j_v}^T \right]^T$ where $\mathbf{g}_j$ is the $j$-th row of $G$. Note that $G^{\mathcal{V}}$ is a $v \times k$ matrix.

In addition, we can represent the decoding rule $g$ by the parity check matrix $H$ instead of the generator matrix $G$ as follows.
\begin{equation}\label{eq:BEC_decoder_LE_H_1}
H^T \mathbf{\widehat{c}} = \left(H^{\mathcal{E}} \right)^T \widehat{\mathbf{c}}^{\mathcal{E}} + \left(H^{\mathcal{V}} \right)^T \widehat{\mathbf{c}}^{\mathcal{V}} = \mathbf{0}
\end{equation}
where $\mathcal{E}=\left\{i_1,\cdots, i_e\right\}$ indicates the locations of $e$ erased bits such that $\mathcal{E} \cup \mathcal{V} = \left\{ 1,2,\ldots, n \right\}$ and $n=e+v$. Note that $\widehat{\mathbf{c}}^{\mathcal{E}}=\left(\widehat{c}_{i_1}, \cdots, \widehat{c}_{i_e}\right)^T$, $\widehat{\mathbf{c}}^{\mathcal{V}}=\left(\widehat{c}_{j_1}, \cdots, \widehat{c}_{j_v}\right)^T$, $H^{\mathcal{E}}=\left[ \mathbf{h}_{i_1}^T, \cdots, \mathbf{h}_{i_e}^T \right]^T$ and $H^{\mathcal{V}}=\left[ \mathbf{h}_{j_1}^T, \cdots, \mathbf{h}_{j_v}^T \right]^T$ where $\mathbf{h}_i$ is the $i$-th row of $H$.

From the channel model of BEC, it is clear that $\widehat{\mathbf{c}}^{\mathcal{V}} = \mathbf{y}^{\mathcal{V}} = \mathbf{c}^{\mathcal{V}}$ and we have to estimate the erased bits of $\mathbf{c}$, i.e., $\widehat{\mathbf{c}}^{\mathcal{E}}$. Thus, \eqref{eq:BEC_decoder_LE_H_1} can be represented by the following linear equations.
\begin{equation}\label{eq:BEC_decoder_LE_H}
\left(H^{\mathcal{E}} \right)^T \widehat{\mathbf{c}}^{\mathcal{E}} = \mathbf{q}
\end{equation}
where $\mathbf{q} = \left(H^{\mathcal{V}} \right)^T \mathbf{c}^{\mathcal{V}}$. $\widehat{\mathbf{m}}$ can be obtained from $\widehat{\mathbf{c}}$. Note that $\left(H^{\mathcal{E}}\right)^T$ is a $(n - k) \times e$ matrix.

Because of the weak law of large numbers, we can claim that $nC_{\mathrm{BEC}} - \epsilon \le v = n - e \le nC_{\mathrm{BEC}} + \epsilon$ with high probability for sufficiently large $n$. Assuming that $R < C_{\mathrm{BEC}} - \epsilon$, we can claim that \eqref{eq:BEC_decoder_LE} and \eqref{eq:BEC_decoder_LE_H} are \emph{overdetermined} because of $v > k$ and $n - k > e$.

Since $\dim \left( \mathcal{C} \right) = k$, there exists exactly one solution of \eqref{eq:BEC_decoder_LE} so long as $\rank \left(G^{\mathcal{V}}\right) = k$. If $\rank \left(G^{\mathcal{V}}\right) < k$, there are several solutions, which may result in decoding failure. Similarly, there exists exactly one solution of \eqref{eq:BEC_decoder_LE_H} so long as $\rank \left( H^{\mathcal{E}}\right) = e$. Otherwise, there are several solutions, which may result in decoding failure.

The following Lemma and its proof have been known in coding theory community.

\begin{lemma}[\cite{Barg2014}] \label{lemma:BEC_UB} The upper bound on the probability of decoding failure of the decoding rule $g$ is given by
\begin{equation}\label{eq:BEC_UB}
P\left(D=0 \mid |\mathcal{E}|=e \right) \le \frac{\sum_{w=d_{\mathrm{min}}}^{e}{A_w \binom{n-w}{e-w}}}{\binom{n}{e}}
\end{equation}
where $A_w$ is the weight distribution of $\mathcal{C}$. Also, $\mathcal{E}$ represents the set of erased locations in the channel output vector $\mathbf{y}$ and $|\mathcal{E}|=e$ is the number of erasures in $\mathbf{y}$.
\end{lemma}

\begin{IEEEproof}
Without loss of generality, we can assume that the all-zero codeword $\mathbf{0}$ has been transmitted and there exists a nonzero codeword $\mathbf{c}$ of Hamming weight $w$ such that $\Psi_w(\mathbf{c}) \subseteq \mathcal{E}$ where $\Psi_w(\mathbf{c})=\left\{ i \mid c_i \ne 0 \right\}$ denotes the locations of nonzero elements of $\mathbf{c}$ and $\mathcal{E}=\left\{i_1, \cdots, i_e \right\}$ denotes the locations of $e$ erasures. From the given decoding rule, $\mathbf{y}$ agrees with two codewords $\mathbf{0}$ and $\mathbf{c}$ on unerased bits, which may result in decoding failure. Meanwhile, if there is no nonzero codeword $\mathbf{c}$ such that $\Psi_w(\mathbf{c}) \subseteq \mathcal{E}$, then $\mathbf{y}$ agrees with only $\mathbf{0}$ on the unerased bits and the decoding succeeds.

For a nonzero $\mathbf{c}$ such that $\Psi_w(\mathbf{c}) \subseteq \mathcal{E}$, the number of possible $\mathcal{E}$ is $\binom{n-w}{e-w}$. Due to double counting, the number of possible $\mathcal{E}$ which results in decoding failure will be less than or equal to $\sum_{w=d_{\mathrm{min}}}^{e}{A_w \binom{n-w}{e-w}}$. Since the number of all possible $\mathcal{E}$ such that $|\mathcal{E}|=e$ is $\binom{n}{e}$, the upper bound on $P\left(D=0 \mid |\mathcal{E}|=e \right)$ is given by \eqref{eq:BEC_UB}.
\end{IEEEproof}

From the upper bound in Lemma~\ref{lemma:BEC_UB}, it is clear that $P\left(D=0 \mid |\mathcal{E}|=e \right) = 0$ for $e < d_{\mathrm{min}}$. The following Lemma shows that $P\left(D=0 \mid |\mathcal{E}|=e \right)$ can be obtained exactly for $d_{\mathrm{min}} \le e \le d_{\mathrm{min}} + \left\lfloor \frac{d_{\mathrm{min}}-1}{2} \right\rfloor$ where $\left\lfloor x \right\rfloor$ represents the largest integer not greater than $x$.

\begin{lemma} \label{lemma:BEC_exact} For $e \le d_{\mathrm{min}} + t$ where $t = \left\lfloor \frac{d_{\mathrm{min}}-1}{2} \right\rfloor$, $P\left( D=0 \mid |\mathcal{E}|=e \right)$ is given by
\begin{equation} \label{eq:BEC_exact}
P\left(D=0 \mid |\mathcal{E}|=e \right) = \frac{1}{2} \cdot \frac{\sum_{w=d_{\mathrm{min}}}^{e}{A_w \binom{n-w}{e-w}}}{\binom{n}{e}}.
\end{equation}
\end{lemma}

\begin{IEEEproof}
Without loss of generality, we can assume that the all-zero codeword $\mathbf{0}$ has been transmitted and suppose that there exists \emph{only one} nonzero codeword $\mathbf{c}$ of Hamming weight $w$ such that $\Psi_w(\mathbf{c}) \subseteq \mathcal{E}$. Since there are only two possible candidates such as $\mathbf{0}$ and $\mathbf{c}$ to guess the transmitted codeword, \eqref{eq:BEC_exact} is true. Thus, we need to show that there exists only one nonzero codeword $\mathbf{c}$ such that $\Psi_w(\mathbf{c}) \subseteq \mathcal{E}$ for $d_{\mathrm{min}} \le e \le d_{\mathrm{min}} + t$.

Suppose that there are two nonzero codewords $\mathbf{c}_1, \mathbf{c}_2 \in \mathcal{C}$ such that $\|\mathbf{c}_1\| = w_1$ and $\|\mathbf{c}_2\| = w_2$ where $d_{\mathrm{min}} \le w_1 \le w_2$. The locations of nonzero elements of $\mathbf{c}_1$ and $\mathbf{c}_2$ are given by
\begin{align}
\Psi_{w_1}\left(\mathbf{c}_1\right)&= \left\{ i_{1,1},\ldots,i_{1, w_1} \right\}, \\
\Psi_{w_2}\left(\mathbf{c}_2\right)&= \left\{ i_{2,1},\ldots,i_{2, w_2} \right\}.
\end{align}
Let $\Psi_{\alpha} = \left\{i_1, \ldots, i_{\alpha} \right\}$ denote $\Psi_{\alpha} = \Psi_{w_1}\left(\mathbf{c}_1\right) \cap \Psi_{w_2}\left(\mathbf{c}_2\right)$. Then $\Psi_{w_1}\left(\mathbf{c}_1\right)$ and $\Psi_{w_2}\left(\mathbf{c}_2\right)$ are given by
\begin{align}
\Psi_{w_1}\left(\mathbf{c}_1\right)&= \Psi_{\alpha} \cup \left\{ i_{1,1}',\ldots,i_{1, \beta_1}' \right\}, \\
\Psi_{w_2}\left(\mathbf{c}_2\right)&= \Psi_{\alpha} \cup \left\{ i_{2,1}',\ldots,i_{2, \beta_2}' \right\}
\end{align}
where $i_{1, j_1}'$ for $j_1 \in \left\{1, \ldots, \beta_1 \right\}$ and $i_{2, j_2}'$ for $j_2 \in \left\{1, \ldots, \beta_2 \right\}$ are the reindexed locations of nonzero elements of $\mathbf{c}_1$ and $\mathbf{c}_2$ that are mutually disjoint with $\Psi_{\alpha}$. Note that $\left\{ i_{1,1}',\ldots,i_{1, \beta_1}' \right\} \cap \left\{ i_{2,1}',\ldots,i_{2, \beta_2}' \right\} = \emptyset $, $\beta_1 = w_1 - \alpha$ and $\beta_2 = w_2 - \alpha$.

Due to the property of linear codes, $\mathbf{c}_3 = \mathbf{c}_1 + \mathbf{c}_2$ is also a codeword of $\mathcal{C}$, i.e., $\mathbf{c}_3 \in \mathcal{C}$ and $\|\mathbf{c}_3\|=\beta_1 + \beta_2$. Also, the following conditions should hold because of the definition of $d_{\mathrm{min}}$.
\begin{align}
\alpha + \beta_1 &\ge d_{\mathrm{min}}, \\
\alpha + \beta_2 &\ge d_{\mathrm{min}}, \\
\beta_1 + \beta_2 &\ge d_{\mathrm{min}}
\end{align}
Thus, we can claim that $2 \left( \alpha + \beta_1 + \beta_2 \right) \ge 3 d_{\mathrm{min}}$, which results in $\alpha + \beta_1 + \beta_2 \ge d_{\mathrm{min}} + \left\lfloor \frac{d_{\mathrm{min}} + 1}{2} \right\rfloor = d_{\mathrm{min}} + t +1 $ since $\alpha + \beta_1 + \beta_2$ is an integer.

If there exist two codewords $\mathbf{c}_1$ and $\mathbf{c}_2$ such that $\Psi_{w_1}\left(\mathbf{c}_1\right) \subseteq \mathcal{E}$ and $\Psi_{w_2}\left(\mathbf{c}_2\right) \subseteq \mathcal{E}$ (i.e., $\Psi_{w_1}\left(\mathbf{c}_1\right) \cup \Psi_{w_2}\left(\mathbf{c}_2\right) \subseteq\mathcal{E}$), it means that $e \ge \alpha + \beta_1 + \beta_2 \ge d_{\mathrm{min}} + t + 1$. Thus, for $d_{\mathrm{min}} \le e \le d_{\mathrm{min}} + \left\lfloor \frac{d_{\mathrm{min}} - 1}{2} \right\rfloor = d_{\mathrm{min}} + t$, there exists at most one nonzero codeword $\mathbf{c}$ such that $\Psi_w(\mathbf{c}) \subseteq \mathcal{E}$.

\end{IEEEproof}

\begin{theorem} \label{thm:BEC_bound} $P\left(D = 0 \mid |\mathcal{E}|=e\right)$ is given by
\begin{numcases}{P\left(D = 0 \mid |\mathcal{E}|=e\right)=}
0 & for $e < d_{\mathrm{min}}$,
\\
\frac{1}{2} \cdot \frac{\sum_{w=d_{\mathrm{min}}}^{e}{A_{w} \binom{n-w}{e-w}}}{\binom{n}{e}}  & for $d_{\mathrm{min}} \le e \le d_{\mathrm{min}}+t$,
\\
\le \frac{\sum_{w=d_{\mathrm{min}}}^{e}{A_{w} \binom{n-w}{e-w}}}{\binom{n}{e}}  & for $ e > d_{\mathrm{min}}+t$.
\end{numcases}
\end{theorem}
\begin{IEEEproof}The proof comes from the definition of $d_{\mathrm{min}}$ in \eqref{eq:BEC_dmin}, Lemma~\ref{lemma:BEC_UB} and Lemma~\ref{lemma:BEC_exact}.
\end{IEEEproof}

\begin{theorem}[\cite{Barg2014}] The decoding rule of $g$ is a capacity achieving scheme. \label{thm:BEC}
\end{theorem}

\begin{IEEEproof}
The decoding failure probability is given by
\begin{align}
P\left(D=0\right)& = P\left(D=0, |\mathcal{E}| \le n(\alpha + \epsilon) \right) + P\left(D=0, |\mathcal{E}| > n(\alpha + \epsilon) \right) \label{eq:BEC_CAS_pf_1}\\
    & \le \sum_{e=1}^{n(\alpha + \epsilon)}{P(D=0, |\mathcal{E}|=e)} + \epsilon' \\
    & \le \sum_{e=1}^{n(\alpha + \epsilon)}{P(D=0 \mid |\mathcal{E}|=e)} + \epsilon' \label{eq:BEC_conditional_prob}\\
    & \le \sum_{e=1}^{n(\alpha + \epsilon)}{\frac{\sum_{w=d_{\mathrm{min}}}^{e}{A_w \binom{n-w}{e-w}}}{\binom{n}{e}}} + \epsilon' \label{eq:BEC_CAS_pf_UB} \\
    & \le \frac{n}{2^{n-k} }\sum_{e=1}^{n(\alpha + \epsilon)}{\frac{\sum_{w=d_{\mathrm{min}}}^{e}{\binom{n}{w} \binom{n-w}{e-w}}}{\binom{n}{e}}} + \epsilon' \label{eq:BEC_CAS_pf_Aw} \\
    & \le \frac{n}{2^{n-k} }\sum_{e=1}^{n(\alpha + \epsilon)}{\frac{\sum_{w=d_{\mathrm{min}}}^{e}{\binom{e}{w} \binom{n}{e}}}{\binom{n}{e}}} + \epsilon' \label{eq:BEC_CAS_binom} \\
    & \le \frac{n}{2^{n-k} }\sum_{e=1}^{n(\alpha + \epsilon)}{\sum_{w=d_{\mathrm{min}}}^{e}{\binom{e}{w}}} + \epsilon' \\
    & \le \frac{n}{2^{n-k} }\sum_{e=1}^{n(\alpha + \epsilon)}{2^{e}} + \epsilon' \\
    & \le n^2 2^{k-n} (\alpha + \epsilon) 2^{n (\alpha + \epsilon)} + \epsilon' \\
    & = n^2 (\alpha + \epsilon) 2^{n \left\{R - (1 - \alpha) + \epsilon \right\}} + \epsilon' \label{eq:BEC_CAS_capacity}
\end{align}
where we assume that $n(\alpha + \epsilon)$ is an integer without loss of generality in \eqref{eq:BEC_CAS_pf_1}. Also, \eqref{eq:BEC_CAS_pf_UB} follows from \eqref{eq:BEC_UB} in Lemma~\ref{lemma:BEC_UB}. \eqref{eq:BEC_CAS_pf_Aw} follows from the fact that there exists an $\left[n, k\right]$ binary linear code whose weight distribution is bounded by \cite{Barg2014}
\begin{equation} \label{eq:Aw_UB}
A_w \le \frac{n}{2^{n-k}} \binom{n}{w}.
\end{equation}
Also, \eqref{eq:BEC_CAS_binom} follows from $\binom{n}{w} \binom{n-w}{e-w} = \binom{e}{w} \binom{n}{e}$.

If $R < 1 - \alpha - \epsilon = C_{\mathrm{BEC}} - \epsilon$ and $n$ is sufficiently large, \eqref{eq:BEC_CAS_capacity} goes to zero. Thus, the decoding rule of $g$ achieves $C_{\mathrm{BEC}}$.
\end{IEEEproof}

\begin{remark} \label{remark:BEC} We can show that the decoding rule $g$ is a capacity achieving scheme without considering the weight distribution of codes. If each element of $G$ in \eqref{eq:BEC_decoder_LE} is selected uniformly at random from $\left\{0, 1\right\}$,
\begin{equation} \label{eq:BEC_CAS_G}
P\left( \rank \left(G^{\mathcal{V}}\right) < k \right) = \frac{2^k}{2^v} = 2^{-n(C_{\mathrm{BEC}} - R)}.
\end{equation}
Similarly, if each element of $H$ in \eqref{eq:BEC_decoder_LE_H} is selected uniformly at random from $\left\{0, 1\right\}$,
\begin{equation} \label{eq:BEC_CAS_H}
P\left( \rank \left(H^{\mathcal{E}}\right) < e \right) = \frac{2^e}{2^{n-k}} = 2^{-n(C_{\mathrm{BEC}} - R)}.
\end{equation}
If $R < C_{\mathrm{BEC}}$ and $n$ is sufficiently large, both \eqref{eq:BEC_CAS_G} and \eqref{eq:BEC_CAS_H} go to zero. Thus, the decoding rule of $g$ can achieve $C_{\mathrm{BEC}}$ by solving either \eqref{eq:BEC_decoder_LE} or \eqref{eq:BEC_decoder_LE_H}.
\end{remark}

\begin{remark} The computational complexity of \eqref{eq:BEC_decoder_LE} is $\mathcal{O}\left(k^3\right)$ where $k = nR$. Also, the computational complexity of \eqref{eq:BEC_decoder_LE_H} is $\mathcal{O}\left(e^3\right)$ where $e = n \alpha$. Though both complexities are eventually $\mathcal{O}\left(n^3\right)$, we can choose one of them to reduce the computational complexity. If $\alpha < 0.5$ and $R > 0.5$, the computational complexity of \eqref{eq:BEC_decoder_LE_H} is less than that of \eqref{eq:BEC_decoder_LE}.
\end{remark}

\subsection{Binary Defect Channel}

We will use the notations of \cite{Heegard1983, Kim2013} with slight modifications for the BDC. We define an additional variable ``$\lambda$'' (denoting the defect-free state) and the channel state $S \in \{0, 1, \lambda\}$. Let ``$\circ$'' denote the operator  $\circ:\{0, 1\} \times \left\{0, 1, \lambda\right\} \rightarrow \{0, 1\}$ by
\begin{equation}\label{eq:circ_operator}
x \circ s =
\begin{cases}
x, & \text{if } s = \lambda ; \\
s, & \text{if } s \ne \lambda.
\end{cases}
\end{equation}
An $n$-cell memory with defects is modeled by
\begin{equation}\label{eq:BDC_channel_model}
\mathbf{y} = \mathbf{x} \circ \mathbf{s}
\end{equation}
where $\mathbf{x} \in \left\{0, 1\right\}^n$ is the channel input vector and $\mathbf{y} \in \left\{0, 1\right\}^n$ is the channel output vector. Also, $\mathbf{s} \in \left\{0, 1, \lambda \right\}^n$ is the channel state vector which has the information of defect locations and stuck-at values. Note that $\circ$ is the vector component-wise operator. The number of defects is equal to the number of non-$\lambda$ components in $\mathbf{s}$. The number of errors due to defects is given by
\begin{equation}\label{eq:BDC_num_errors}
\| \mathbf{x} \circ \mathbf{s} - \mathbf{x} \|.
\end{equation}

As shown in Fig.~\ref{fig:BDC},
\begin{equation}\label{eq:BDC_random_channel_model}
\begin{aligned}
P(S=s)& =
\begin{cases}
1-\beta,  & s = \lambda; \\
\frac{\beta}{2}, &  s = 0; \\
\frac{\beta}{2}, &  s = 1.
\end{cases}
\end{aligned}
\end{equation}

In \cite{Heegard1983}, Heegard discussed additive encoding and defined the $[n,k,l]$ partitioned linear block code (PLBC) which consists of a pair of linear subspaces $\mathcal{C}_1 \subset \{0, 1\}^n$ and $\mathcal{C}_0 \subset \{0, 1\}^n$ of dimension $k$ and $l$ such that $\mathcal{C}_1 \cap \mathcal{C}_0 =\{ \mathbf{0}\}$. Then the direct sum is given by
\begin{equation}\label{eq:direct_sum}
\mathcal{C} \triangleq \mathcal{C}_1 + \mathcal{C}_0 = \{ \mathbf{c} = \mathbf{c}_1 + \mathbf{c}_0 | \mathbf{c}_1 \in \mathcal{C}_1 , \mathbf{c}_0 \in \mathcal{C}_0 \}.
\end{equation}

\emph{Encoding:} A message $\mathbf{m} \in \{0, 1\}^k$ is encoded to a corresponding codeword $\mathbf{c}$ as follows.
\begin{equation}
\mathbf{c}= \mathbf{c}_1 + \mathbf{c}_0 = G_1 \mathbf{m} + G_0 \mathbf{d}
\end{equation}
where $\mathbf{c}_1 = G_1 \mathbf{m}$ and $\mathbf{c}_0 = G_0 \mathbf{d}$. The generator matrix for $\mathbf{c}_1$ is $G_1 = \left[ I_k \quad 0_{k,l} \right]^T$ where $I_k$ is the $k$-dimensional identity matrix and $0_{k,l}$ is the zero matrix with size of $k \times l$. Also, the generator matrix for $\mathbf{c}_0$ is $G_0$ which is an $n \times l$ matrix. Note that $k + l = n$.

Since the channel state vector $\mathbf{s}$ is available at the encoder, the encoder should choose $\mathbf{d} \in \{0, 1\}^l$ judiciously. The optimal parity $\mathbf{d}$ is chosen to minimize the number of errors due to defects, i.e., $ \|(\mathbf{c} \circ \mathbf{s}) - \mathbf{c} \|$.

\emph{Decoding:} The decoder estimates the message $\mathbf{m}$ as follows.
\begin{equation} \label{eq:BDC_decoding}
\widehat{\mathbf{m}} = \widetilde{G}_1^T \mathbf{y} = H_0^T \mathbf{y}
\end{equation}
where $\widehat{\mathbf{m}}$ is the estimate of $\mathbf{m}$ and the channel output vector $\mathbf{y} = \mathbf{c} \circ \mathbf{s}$ is given by \eqref{eq:BDC_channel_model}. The message inverse matrix $\widetilde{G}̃_1$ is defined as an $n \times k$ matrix such that $\widetilde{G}̃_1^T G_1=I_k$, and $\widetilde{G}̃_1^T G_0 =0_{k,l}$ \cite{Heegard1983}. For the BDC, the message inverse matrix $\widetilde{G}̃_1$ defined by Heegard will be the systematic parity check matrix $H_0$ since it satisfies two conditions for the message inverse matrix.

For convenience, we will define a random variable $M$ as follows.
\begin{equation}
M =
\begin{cases}
0, & \|(\mathbf{c} \circ \mathbf{s}) - \mathbf{c} \| \ne 0 \text{ (masking failure)}; \\
1, & \|(\mathbf{c} \circ \mathbf{s}) - \mathbf{c} \| = 0 \text{ (masking success)}
\end{cases}
\end{equation}

The minimum distance $d_0$ of an $[n,k,l]$ PLBC is given by \cite{Tsybakov1975b, Heegard1983}
\begin{align} \label{eq:BDC_dmin}
d_0 &= \underset{
\substack{
\mathbf{c} \ne \mathbf{0} \\
G_0^T \mathbf{c}= \mathbf{0}
}}
{\text{min }} \|\mathbf{c}\|
\end{align}
which means that any $d_0 - 1$ rows of $G_0$ are linearly independent. If $u < d_0$, all $u$ defects will be masked and $\|(\mathbf{c} \circ \mathbf{s})-\mathbf{c} \|=0$ (i.e., $M = 1$), which will be shown in Lemma~\ref{lemma:BDC_UB}. 

The encoder knows the channel state vector $\mathbf{s}$ and tries to minimize $ \|(\mathbf{c} \circ \mathbf{s}) - \mathbf{c} \|$ by choosing $\mathbf{d}$ judiciously. The encoding of PLBC includes an implicit optimization problem which can be formulated as follows \cite{Heegard1983, Lastras-Montano2010, Hwang2011a}.
\begin{align}
\mathbf{d}^*  & =  \underset{\mathbf{d}}{\text{argmin }} \left\|G_0^{\mathcal{U}} \mathbf{d} +  G_1^{\mathcal{U}} \mathbf{m} - \mathbf{s}^{\mathcal{U}} \right\|  \\
&= \underset{\mathbf{d}}{\text{argmin }} \left\| G_0^{\mathcal{U}} \mathbf{d} + \mathbf{b}^{\mathcal{U}} \right\| \label{eq:BDC_opt_problem}
\end{align}
where $\mathcal{U}=\left\{i_1,\cdots,i_u \right\}$ indicates the set of locations of $u$ defects and $\mathbf{b} = G_1 \mathbf{m} - \mathbf{s}$. Thus, $\mathbf{b}^{\mathcal{U}}$ is given by
\begin{equation}\label{eq:BDC_b}
\mathbf{b}^{\mathcal{U}} = G_1^{\mathcal{U}} \mathbf{m} - \mathbf{s}^{\mathcal{U}}
\end{equation}
where $s^{\mathcal{U}}=\left(s_{i_1},\cdots,s_{i_u}\right)^T$, $G_0^{\mathcal{U}}=\left[\mathbf{g}_{0,i_1}^T,\cdots,\mathbf{g}_{0,i_u}^T \right]^T$, and $G_1^{\mathcal{U}}=\left[\mathbf{g}_{1,i_1}^T,\cdots,\mathbf{g}_{1,i_u}^T \right]^T$. Note that $\mathbf{g}_{0,i}$ and $\mathbf{g}_{1,i}$ are the $i$-th rows of $G_0$ and $G_1$ respectively. Also, $\left\|G_0^{\mathcal{U}} \mathbf{d}+ \mathbf{b}^{\mathcal{U}} \right\|$ represents the number of errors due to defects which is equivalent to \eqref{eq:BDC_num_errors}.

By solving the optimization problem of \eqref{eq:BDC_opt_problem}, the number of errors due to defects will be minimized. However, the computational complexity for solving \eqref{eq:BDC_opt_problem} is exponential, which is impractical \cite{Hwang2011a}.

Instead of solving the impractical optimization problem, $C_{\mathrm{BDC}}$ can be achieved by solving the following system of linear equations \cite{Jagmohan2010a}.
\begin{equation}\label{eq:BDC_encoder_LE}
G_0^{\mathcal{U}} \mathbf{d} = \mathbf{b}^{\mathcal{U}}
\end{equation}
Because of the weak law of large numbers, we can claim that $n C_{\mathrm{BDC}} - \epsilon \le n - u \le n C_{\mathrm{BDC}} + \epsilon$ with high probability for sufficiently large $n$. For $R < C_{\mathrm{BDC}} - \epsilon$, \eqref{eq:BDC_encoder_LE} is \emph{underdetermined} since $G_0^{\mathcal{U}}$ is a $u \times (n - k)$ matrix. If \eqref{eq:BDC_encoder_LE} has at least one solution, the masking succeeds since $\left\| G_0^{\mathcal{U}} \mathbf{d} + \mathbf{b}^{\mathcal{U}} \right\|=0$.

In \cite{Kim2013}, an upper bound on the probability of masking failure was derived and numerical results showed that the upper bound is tight and the performance of partitioned Bose, Chaudhuri, Hocquenghem (PBCH) codes is not far from $C_{\textrm{BDC}}$. We will show that the additive encoding achieves $C_{\mathrm{BDC}}$ by using the upper bound in \cite{Kim2013}, which explains why the performance of PBCH codes is good. The PBCH code is a special class of PLBC and its generator matrices and minimum distances can be designed by a similar method such as standard BCH codes \cite{Heegard1983, Kim2013:plbc}

First, we will present the upper bound on the probability of masking failure for $u$ defects through the following Lemma~\ref{lemma:BDC_lower_upper_bounds},~\ref{lemma:BDC_UB},~\ref{lemma:BDC_exact}, and Theorem~\ref{thm:BDC_bound}.

\begin{lemma}[{\cite{Kim2013}}] \label{lemma:BDC_lower_upper_bounds} The lower and upper bounds on $P(M=0||\mathcal{U}|=u)$ is given by
\begin{equation} \label{eq:BDC_lower_upper_bounds}
\begin{aligned}
\frac{1}{2} \cdot P \left( \rank \left( G_0^{\mathcal{U}} \right) < u \mid |\mathcal{U}|=u \right) &\le P\left(M=0 \mid |\mathcal{U}|=u\right) \\
 & \le P \left( \rank \left( G_0^{\mathcal{U}} \right) < u \mid |\mathcal{U}|=u \right).
\end{aligned}
\end{equation}
\end{lemma}

\begin{IEEEproof}
\eqref{eq:BDC_encoder_LE} has at least one solution if and only if
\begin{equation} \label{eq:BDC_LE_sol_exist}
\rank \left( G_0^{\mathcal{U}} \right) = \rank \left( G_0^{\mathcal{U}} \mid \mathbf{b}^{\mathcal{U}} \right)
\end{equation}
where $\left( G_0^{\mathcal{U}} \mid \mathbf{b}^{\mathcal{U}} \right)$ is the augmented matrix.

If $\rank \left( G_0^{\mathcal{U}} \right) = u$, \eqref{eq:BDC_encoder_LE} has at least one solution since \eqref{eq:BDC_LE_sol_exist} holds. Thus, $P\left(M=0 \mid |\mathcal{U}|=u\right)=0$.

If $\rank \left( G_0^{\mathcal{U}} \right) = u - j$ for $1 \le j \le u$, the last $j$ rows of the row reduced echelon form of $G_0^{\mathcal{U}}$ are zero vectors. In order to satisfy the condition of \eqref{eq:BDC_LE_sol_exist}, the last $j$ elements of the column vector $\mathbf{b}^{\mathcal{U}}$ should also be zeros. The probability that the last $j$ elements of the column vector $\mathbf{b}^{\mathcal{U}}$ are zeros is $\frac{1}{2^j}$ since $P(S=0 \mid S \ne \lambda) = P(S=1 \mid S \ne \lambda) = \frac{1}{2}$. Thus, $P\left(M=0 \mid |\mathcal{U}|=u\right)$ is given by
\begin{equation} \label{eq:BDC_exact}
P\left(M=0 \mid |\mathcal{U}|=u\right) = \sum_{j=1}^{u}{\frac{2^j-1}{2^j}P \left( \rank \left( G_0^{\mathcal{U}} \right) =u - j \mid |\mathcal{U}|=u \right)}
\end{equation}
which results in \eqref{eq:BDC_lower_upper_bounds}.
\end{IEEEproof}

\begin{lemma}[\cite{Kim2013}] \label{lemma:BDC_UB} The upper bound on $P(M=0||\mathcal{U}|=u)$ is given by
\begin{equation}\label{eq:BDC_UB}
P\left(M=0 \mid |\mathcal{U}|=u \right) \le \frac{\sum_{w=d_{0}}^{u}{B_{w} \binom{n-w}{u-w}}}{\binom{n}{u}}
\end{equation}
where $B_{w}$ is the weight distribution of $\mathcal{C}_{0}^{\perp}$ (i.e., the dual code of $\mathcal{C}_0$).
\end{lemma}

\begin{IEEEproof}
Suppose that there exists a nonzero codeword $\mathbf{c}^{\perp} \in \mathcal{C}_{0}^{\perp}$ of Hamming weight $w$. Note that $G_0$ is the parity check matrix of $\mathcal{C}_{0}^{\perp}$. Let $\Psi_w(\mathbf{c}^{\perp})=\left\{ i \mid c_i^{\perp} \ne 0 \right\}$ denote the locations of nonzero elements of $\mathbf{c}^{\perp}$ and $\mathcal{U} =\left\{i_1,\ldots,i_u \right\}$ denote the locations of $u$ defects.

If $\Psi_w(\mathbf{c}^{\perp}) \subseteq \mathcal{U}$, $\rank \left( G_0^{\mathcal{U}} \right) < u$. The reason is that $G_0^{\Psi_{w}(\mathbf{c}^{\perp})}$ is a submatrix of $G_0^{\mathcal{U}}$ and the rows of $G_0^{\Psi_{w}(\mathbf{c}^{\perp})}$ are linearly dependent since $G_0^T \mathbf{c}^{\perp} = \mathbf{0}$.

For any $\mathbf{c}^{\perp}$ such that $\Psi_w(\mathbf{c}^{\perp}) \subseteq \mathcal{U}$, the number of possible $\mathcal{U}$ is $\binom{n-w}{u-w}$. Due to double counting, the number of $\mathcal{U}$ which results in $\rank \left( G_0^{\mathcal{U}} \right) < u$  will be less than or equal to $\sum_{w=d_{0}}^{u}{B_w \binom{n-w}{u-w}}$. Since the number of all possible $\mathcal{U}$ such that $|\mathcal{U}|=u$ is $\binom{n}{u}$,
\begin{equation} \label{eq:BDC_upper_bound_rank}
P\left(\rank \left( G_0^{\mathcal{U}} \right) < u \mid |\mathcal{U}|=u \right) \le \frac{\sum_{w=d_{0}}^{u}{B_{w} \binom{n-w}{u-w}}}{\binom{n}{u}}.
\end{equation}

By \eqref{eq:BDC_lower_upper_bounds} and \eqref{eq:BDC_upper_bound_rank}, the upper bound on $P\left(M=0 \mid |\mathcal{U}|=u \right)$ is given by \eqref{eq:BDC_UB}.
\end{IEEEproof}

From the upper bound in Lemma~\ref{lemma:BDC_UB}, it is clear that $P\left(M=0 \mid |\mathcal{U}|=u \right) = 0$ for $u < d_{0}$. It is worth mentioning that the upper bound on $P\left(M=0 \mid |\mathcal{U}|=u\right)$ for the BDC is similar to the upper bound on $P\left(D=0 \mid |\mathcal{E}|=e\right)$ for the BEC presented in Lemma~\ref{lemma:BEC_UB}.

Similar to Lemma~\ref{lemma:BEC_exact}, the following Lemma shows that $P\left(M=0 \mid |\mathcal{U}|=u \right)$ can be obtained exactly for $d_0 \le u \le d_0 + \left\lfloor \frac{d_0-1}{2} \right\rfloor$.

\begin{lemma}[\cite{Kim2013}] \label{lemma:BDC_exact}For $u \le d_0 + t_0$ where $t_0 = \left\lfloor \frac{d_0 - 1}{2} \right\rfloor$, $P\left(M=0 \mid |\mathcal{U}|=u \right)$ is given by
\begin{equation}
\label{eq:BDC_exact_condition}
P\left(M=0 \mid |\mathcal{U}|=u \right) = \frac{1}{2} \cdot \frac{\sum_{w=d_{0}}^{u}{B_{w} \binom{n-w}{u-w}}}{\binom{n}{u}}.
\end{equation}
\end{lemma}

\begin{IEEEproof}
The proof has two parts. First, we will show that
\begin{equation} \label{eq:BDC_exact_condition_pf1}
P\left(\rank \left( G_0^{\mathcal{U}} \right) < u \mid |\mathcal{U}|=u \right) = \frac{\sum_{w=d_{0}}^{u}{B_{w} \binom{n-w}{u-w}}}{\binom{n}{u}}
\end{equation}
for $u \le d_0 + t_0$, which means that there is no double counting in \eqref{eq:BDC_upper_bound_rank}. Second, we will prove that
\begin{equation} \label{eq:BDC_exact_condition_pf2}
P\left(\rank \left( G_0^{\mathcal{U}} \right) < u \mid |\mathcal{U}|=u \right) = P\left(\rank \left( G_0^{\mathcal{U}} \right) = u-1 \mid |\mathcal{U}|=u \right)
\end{equation}
for $u \le d_0 + t_0$, which means that $P\left(\rank \left( G_0^{\mathcal{U}} \right) \le u-2 \mid |\mathcal{U}|=u \right)=0$.

Then, $P\left(M=0 \mid |\mathcal{U}|=u \right)$ is given by
\begin{align}
P\left(M=0 \mid |\mathcal{U}|=u \right) &= \frac{1}{2} \cdot P \left( \rank \left( G_0^{\mathcal{U}} \right) = u - 1 \mid |\mathcal{U}|=u \right) \label{eq:BDC_exact_condition_step1}\\
    &= \frac{1}{2} \cdot \frac{\sum_{w=d_{0}}^{u}{B_{w} \binom{n-w}{u-w}}}{\binom{n}{u}} \label{eq:BDC_exact_condition_step2}
\end{align}
where \eqref{eq:BDC_exact_condition_step1} follows from \eqref{eq:BDC_exact} and \eqref{eq:BDC_exact_condition_pf2}. Also, \eqref{eq:BDC_exact_condition_step2} follows from \eqref{eq:BDC_exact_condition_pf1}.

1) Proof of \eqref{eq:BDC_exact_condition_pf1}

Suppose that there are two nonzero codewords $\mathbf{c}_1^{\perp}, \mathbf{c}_2^{\perp} \in \mathcal{C}_0^{\perp}$ such that $\|\mathbf{c}_1^{\perp}\| = w_1$ and $\|\mathbf{c}_2^{\perp}\| = w_2$. Without loss of generality, we can assume that $d_0 \le w_1 \le w_2$. The locations of nonzero elements of $\mathbf{c}_1^{\perp}$ and $\mathbf{c}_2^{\perp}$ are given by
\begin{align}
\Psi_{w_1}\left(\mathbf{c}_1^{\perp}\right)&= \left\{ i_{1,1},\ldots,i_{1, w_1} \right\} \\
\Psi_{w_2}\left(\mathbf{c}_2^{\perp}\right)&= \left\{ i_{2,1},\ldots,i_{2, w_2} \right\}.
\end{align}

Let $\Psi_{\alpha}=\left\{i_1, \ldots, i_{\alpha} \right\}$ denote $\Psi_{\alpha} = \Psi_{w_1}\left(\mathbf{c}_1^{\perp}\right) \cap \Psi_{w_2}\left(\mathbf{c}_2^{\perp}\right)$. Then $\Psi_{w_1}\left(\mathbf{c}_1^{\perp}\right)$ and $\Psi_{w_2}\left(\mathbf{c}_2^{\perp}\right)$ are given by
\begin{align}
\Psi_{w_1}\left(\mathbf{c}_1^{\perp}\right)&= \Psi_{\alpha} \cup \left\{ i_{1,1}',\ldots,i_{1, \beta_1}' \right\} \\
\Psi_{w_2}\left(\mathbf{c}_2^{\perp}\right)&= \Psi_{\alpha} \cup \left\{ i_{2,1}',\ldots,i_{2, \beta_2}' \right\}
\end{align}
where $i_{1, j_1}'$ for $j_1 \in \left\{1, \ldots, \beta_1 \right\}$ and $i_{2, j_2}'$ for $j_2 \in \left\{1, \ldots, \beta_2 \right\}$ are the reindexed locations of nonzero elements of $\mathbf{c}_1^{\perp}$ and $\mathbf{c}_2^{\perp}$ that are mutually disjoint with $\Psi_{\alpha}$. Note that $\left\{ i_{1,1}',\ldots,i_{1, \beta_1}' \right\} \cap \left\{ i_{2,1}',\ldots,i_{2, \beta_2}' \right\} = \emptyset $, $\beta_1 = w_1 - \alpha$ and $\beta_2 = w_2 - \alpha$.

Due to the property of linear codes, $\mathbf{c}_3^{\perp} = \mathbf{c}_1^{\perp} + \mathbf{c}_2^{\perp}$ is also a codeword of $\mathcal{C}_0^{\perp}$, i.e., $\mathbf{c}_3^{\perp} \in \mathcal{C}_0^{\perp}$ and $\|\mathbf{c}_3^{\perp} \|=\beta_1 + \beta_2$. Also, the following conditions should hold because of the definition of $d_0$.
\begin{align}
\alpha + \beta_1 &\ge d_0 \\
\alpha + \beta_2 &\ge d_0 \\
\beta_1 + \beta_2 &\ge d_0
\end{align}
Thus, we can claim that $2 \left( \alpha + \beta_1 + \beta_2 \right) \ge 3 d_0$, which results in $\alpha + \beta_1 + \beta_2 \ge d_0 + \left\lfloor \frac{d_0 + 1}{2} \right\rfloor = d_0 + t_0 + 1$ since $\alpha + \beta_1 + \beta_2$ is an integer.

For double counting in \eqref{eq:BDC_upper_bound_rank}, there should exist at least two codewords $\mathbf{c}_1^{\perp}$ and $\mathbf{c}_2^{\perp}$ such that $\Psi_{w_1}\left(\mathbf{c}_1^{\perp}\right) \cup \Psi_{w_2}\left(\mathbf{c}_2^{\perp}\right) \subseteq\mathcal{U}$. It means that double counting occurs only if $u \ge \alpha + \beta_1 + \beta_2 \ge d_0 + t_0 + 1$. Thus, there is no double counting for $u \le d_0 + t_0$. For $u \le d_0 + t_0$, there exists at most one codeword $\mathbf{c}^{\perp}$ such that $\Psi_{w}\left( \mathbf{c}^{\perp} \right) \subseteq \mathcal{U}$.

2) Proof of \eqref{eq:BDC_exact_condition_pf2}

It is clear that $\rank \left( G_0^{\mathcal{U}}\right) = u - 1$ if and only if there exists only one nonzero codeword $\mathbf{c}^{\perp}$ such that $\Psi_{w}\left( \mathbf{c}^{\perp} \right) \subseteq \mathcal{U}$. Note that $\rank \left( G_0^{\mathcal{U}}\right) < u - 1$ if and only if $\mathcal{U}$ includes the locations of nonzero elements of at least two nonzero codewords. We have already shown that there exists at most one nonzero codeword $\mathbf{c}^{\perp}$ such that $\Psi_{w}\left( \mathbf{c}^{\perp} \right) \subseteq \mathcal{U}$ for $u \le d_0 + t_0$.

\end{IEEEproof}

Similar to the upper bound on $P\left( D = 0 \mid |\mathcal{E}|=e \right)$ in Theorem~\ref{thm:BEC_bound} for the BEC, we can provide the upper bound on $P\left( M = 0 \mid |\mathcal{U}|=u \right)$ for the BDC as follows.

\begin{theorem}[\cite{Kim2013}] \label{thm:BDC_bound} $P\left(M = 0 \mid |\mathcal{U}|=u\right)$ is given by
\begin{numcases}{P\left(M = 0 \mid |\mathcal{U}|=u\right)=}
0 & for $u < d_0$,
\\
\frac{1}{2} \cdot \frac{\sum_{w=d_{0}}^{u}{B_{w} \binom{n-w}{u-w}}}{\binom{n}{u}}  & for $d_0 \le u \le d_0+t_0$, \label{eq:BDC_exact_condition_thm}
\\
\le \frac{\sum_{w=d_{0}}^{u}{B_{w} \binom{n-w}{u-w}}}{\binom{n}{u}}  & for $ u > d_0+t_0$.
\end{numcases}
\end{theorem}

\begin{IEEEproof}The proof comes from the definition of $d_0$ in \eqref{eq:BDC_dmin}, Lemma~\ref{lemma:BDC_UB} and Lemma~\ref{lemma:BDC_exact}.
\end{IEEEproof}

Comparing the upper bound on $P\left( D = 0 \mid |\mathcal{E}|=e \right)$ for the BEC and $P\left( M = 0 \mid |\mathcal{U}|=u \right)$ for BDC, the duality of erasures and defects can be seen. The expressions for both upper bounds in Theorem~\ref{thm:BEC_bound} and Theorem~\ref{thm:BDC_bound} are very similar. The one difference is in the minimum distances such as $d_{\textrm{min}}$ and $d_{0}$. For the definition of $d_{\textrm{min}}$, $H$ is the parity check matrix in \eqref{eq:BEC_dmin}. Meanwhile, $G_0$ is the parity check matrix in \eqref{eq:BDC_dmin}. The other difference comes from the weight distributions such as $A_w$ and $B_w$. Note that $A_w$ is the weight distribution of $\mathcal{C}$ and $B_w$ the weight distribution of $\mathcal{C}_0^{\perp}$.

The following Theorem shows that the capacity of the BDC can be achieved by an encoding scheme based on solving the linear equations \eqref{eq:BDC_encoder_LE}.

\begin{theorem} The encoding scheme of solving the linear equations \eqref{eq:BDC_encoder_LE} is a capacity achieving scheme. \label{thm:BDC}
\end{theorem}

\begin{IEEEproof}
The masking failure probability is given by
\begin{align}
P\left( M=0 \right) & = P\left(M=0, U \le n(\beta + \epsilon) \right) + P\left(M=0, U > n(\beta + \epsilon) \right) \label{eq:BDC_CAS_pf_1}\\
    & \le \sum_{u=1}^{n(\beta + \epsilon)}{P(M=0, |\mathcal{U}|=u)} + \epsilon' \\
    & \le \sum_{u=1}^{n(\beta + \epsilon)}{P(M=0 \mid |\mathcal{U}|=u)} + \epsilon' \\
    & \le \sum_{u=1}^{n(\beta + \epsilon)}{ \frac{\sum_{w=d_{0}}^{u}{B_{w} \binom{n-w}{u-w}}}{\binom{n}{u}}} + \epsilon' \label{eq:BDC_CAS_pf_UB} \\
    & \le \frac{n}{2^{n-k} }\sum_{u=1}^{n(\beta + \epsilon)}{\frac{\sum_{w=d_{0}}^{u}{\binom{n}{w} \binom{n-w}{u-w}}}{\binom{n}{u}}} + \epsilon' \label{eq:BDC_CAS_pf_Bw} \\
    & \le \frac{n}{2^{n-k} }\sum_{u=1}^{n(\beta + \epsilon)}{\frac{\sum_{w=d_{0}}^{u}{\binom{u}{w} \binom{n}{u}}}{\binom{n}{u}}} + \epsilon' \label{eq:BDC_CAS_binom} \\
    & \le \frac{n}{2^{n-k} }\sum_{u=1}^{n(\beta + \epsilon)}{\sum_{w=d_{0}}^{u}{\binom{u}{w}}} + \epsilon' \\
    & \le \frac{n}{2^{n-k} }\sum_{u=1}^{n(\beta + \epsilon)}{2^{u}} + \epsilon'\\
    & \le n^2 2^{k-n} (\beta + \epsilon) 2^{n (\beta + \epsilon)} + \epsilon' \\
    & = n^2 (\beta + \epsilon) 2^{n\left\{R - \left(1 - \beta \right) + \epsilon\right\}} + \epsilon' \label{eq:BDC_CAS_capacity}
\end{align}
where we assume that $n(\beta + \epsilon)$ is an integer without loss of generality in \eqref{eq:BDC_CAS_pf_1}. Also, \eqref{eq:BDC_CAS_pf_UB} follows from Lemma~\ref{lemma:BDC_UB}. \eqref{eq:BDC_CAS_pf_Bw} follows from the fact that $\mathcal{C}_{0}^{\perp}$ is an $[n, k]$ linear code and the upper bound on the weight distribution of $\mathcal{C}_{0}^{\perp}$ can be obtained from \eqref{eq:Aw_UB}. \eqref{eq:BDC_CAS_binom} follows from the fact that $\binom{n}{w} \binom{n-w}{u-w} = \binom{u}{w} \binom{n}{u}$.

If $R < 1 - \beta - \epsilon = C_{\mathrm{BDC}} - \epsilon$ and $n$ is sufficiently large, \eqref{eq:BDC_CAS_capacity} goes to zero. Thus, the additive encoding with solving the system of linear equations of \eqref{eq:BDC_encoder_LE} achieves the channel capacity of BDC.
\end{IEEEproof}

It is well known that random binning is a capacity achieving scheme for the BDC. The encoding of random binning is as follows \cite{Heegard1983:capacity, ElGamal2011}: Randomly partition the $2^n$ sequences into $2^{nR}$ equal size subsets (or bins) and associate a different message with each bin. When the $i$-th message is to be stored, search the $i$-th bin for a sequence (or codeword) $\mathbf{c}$ such that $\mathbf{c} \circ \mathbf{s} = \mathbf{c}$. The decoding is to choose the index of the bin that the channel output vector $\mathbf{y}$ belongs to.

The encoding of random binning can be described by the following linear equations \cite{Jagmohan2010a}.
\begin{equation} \label{eq:BDC_random_binning_1}
H_0^T \mathbf{c} = \mathbf{m}
\end{equation}
where $\mathbf{c}$ will be chosen to satisfy $\mathbf{c} \circ \mathbf{s} = \mathbf{c}$. We can see that the linear equations for the encoding of random binning is equivalent to \eqref{eq:BDC_decoding} which represents the decoding of additive encoding, which shows \emph{the duality between additive encoding and random binning}.

\eqref{eq:BDC_random_binning_1} can be modified into
\begin{align}\label{eq:BDC_random_binning_2}
H_0^T \mathbf{c} & = \left(H_0^{\mathcal{U}} \right)^T \mathbf{c}^{\mathcal{U}} + \left(H_0^{\mathcal{W}} \right)^T \mathbf{c}^{\mathcal{W}} \\
    & = \mathbf{m}
\end{align}
where $\mathcal{U}=\left\{i_1,\cdots, i_u\right\}$ indicates the locations of stuck-at defects and $\mathcal{W} = \left\{j_1,\cdots, j_w\right\}$ represents the locations of normal cells such that $\mathcal{U} \cup \mathcal{W} = \left\{ 1,2,\ldots, n \right\}$. Note that $\mathbf{c}^{\mathcal{U}}=\left(c_{i_1}, \cdots, c_{i_u}\right)^T$, $\mathbf{c}^{\mathcal{W}}=\left(c_{j_1}, \cdots, c_{j_w}\right)^T$, $H_0^{\mathcal{U}}=\left[ \mathbf{h}_{0, i_1}^T, \cdots, \mathbf{h}_{0, i_e}^T \right]^T$ and $H_0^{\mathcal{W}}=\left[ \mathbf{h}_{0, j_1}^T, \cdots, \mathbf{h}_{0, j_w}^T \right]^T$ where $\mathbf{h}_{0, i}$ is the $i$-th row of $H_0$. Since $\mathbf{s}^{\mathcal{U}}$ is known to the encoder, the encoder of random binning can set $\mathbf{c}^{\mathcal{U}} = \mathbf{s}^{\mathcal{U}}$. Thus, the random binning can be described by
\begin{equation}\label{eq:BDC_encoder_LE_binning}
\left(H_0^{\mathcal{W}} \right)^T \mathbf{c}^{\mathcal{W}} = \mathbf{m}'
\end{equation}
where $\mathbf{m}' = \mathbf{m} - \left(H_0^{\mathcal{U}} \right)^T \mathbf{s}^{\mathcal{U}}$. The solution of \eqref{eq:BDC_encoder_LE_binning} represents the codeword elements of normal cells. Note that $\left(H_0^{\mathcal{W}} \right)^T$ is a $k \times (n - u)$ matrix. Thus, \eqref{eq:BDC_encoder_LE_binning} is also \emph{underdetermined} for $R < C_{\mathrm{BDC}} - \epsilon$.

\begin{remark} \label{remark:BDC} We can show that both additive encoding and random binning are capacity achieving scheme by the same method in Remark~\ref{remark:BEC}. If each element of $G_0$ in \eqref{eq:BDC_encoder_LE} is selected uniformly at random from $\left\{0, 1\right\}$,
\begin{equation} \label{eq:BDC_CAS_G}
P\left( \rank \left(G_0^{\mathcal{U}}\right) < u \right) = \frac{2^u}{2^{n-k}} = 2^{-n(C_{\mathrm{BDC}} - R)}.
\end{equation}
Similarly, if each element of $H_0$ in \eqref{eq:BDC_encoder_LE_binning} is selected uniformly at random from $\left\{0, 1\right\}$,
\begin{equation} \label{eq:BDC_CAS_H}
P\left( \rank \left(H_0^{\mathcal{W}}\right) < u \right) = \frac{2^k}{2^{n-u}} = 2^{-n(C_{\mathrm{BDC}} - R)}.
\end{equation}
If $R < C_{\mathrm{BDC}}$ and $n$ is sufficiently large, both \eqref{eq:BDC_CAS_G} and \eqref{eq:BDC_CAS_H} go to zero. Thus, both additive encoding and random binning achieve $C_{\mathrm{BDC}}$ by solving linear equations.
\end{remark}

\begin{remark} The computational complexity of additive encoding of \eqref{eq:BDC_encoder_LE} is $\mathcal{O}\left(u^3\right)$ where $u = n\beta$. Also, the computational complexity of random binning of \eqref{eq:BDC_encoder_LE_binning} is $\mathcal{O}\left(k^3\right)$ where $k = n R$. Though both complexities are $\mathcal{O}\left(n^3\right)$, we can claim that \emph{additive encoding is better than random binning} since $\beta$ is generally very small for storage systems, i.e., $u \ll k$.
\end{remark}

\subsection{Duality between Erasures and Defects}

We will discuss the duality of erasures and defects which is summarized in Table~\ref{tab:duality}. In the BEC used for \emph{communication}, the channel input $X \in \left\{0, 1 \right\}$ is binary and the channel output $Y = \left\{ 0, 1, \varepsilon \right\}$ is ternary where the erasure $\varepsilon$ is \emph{neither} 0 nor 1. In the BDC used for \emph{storage}, the channel state $S \in \left\{0, 1, \lambda \right\}$ is ternary whereas the channel input and output are binary. The ternary channel state $S$ informs whether the given cells are stuck-at defects or normal cells. The stuck-at value is \emph{either} 0 or 1.

The expressions for capacities of both channels are quite similar as shown in \eqref{eq:BEC_capacity} and \eqref{eq:BDC_capacity}. In the BEC, the \emph{decoder} corrects erasures by using the information of locations of erasures, whereas the \emph{encoder} masks the defects by using the information of defect locations and stuck-at values in the BDC.

The capacity achieving scheme of the BEC can be represented by the linear equations based on the generator matrix $G$ of \eqref{eq:BEC_decoder_LE} or the linear equations based on the parity check matrix $H$ of \eqref{eq:BEC_decoder_LE_H}. Both linear equations are \emph{overdetermined}. The solution of the linear equations based on $G$ is the \emph{estimate of message} $\widehat{\mathbf{m}}$ and there should be only one $\widehat{\mathbf{m}}$ for decoding success. Also, the solution of the linear equations based on $H$ is the \emph{estimate of erased bits} $\widehat{\mathbf{c}}^{\mathcal{E}}$ which should be only one $\widehat{\mathbf{c}}^{\mathcal{E}}$ for decoding success.

On the other hand, the capacity achieving scheme of the BDC can be described by \emph{underdetermined} linear equations. The \emph{additive encoding} can be represented by the linear equations based on the generator matrix $G_0$ of \eqref{eq:BDC_encoder_LE} whose solution is the \emph{parity} $\mathbf{d}$. Also, the \emph{random binning} can be represented by the linear equations based on the parity check matrix $H_0$ of \eqref{eq:BDC_encoder_LE_binning} whose solution is the \emph{codeword elements of normal cells} $\mathbf{c}^{\mathcal{W}}$. Unlike the coding scheme of the BEC, there can be several solutions of $\mathbf{d}$ or $\mathbf{c}^{\mathcal{W}}$ that matches all stuck-at defects.

We can see the duality between erasures and defects by comparing the solution $\widehat{\mathbf{m}}$ of \eqref{eq:BEC_decoder_LE} and the solution $\mathbf{d}$ of \eqref{eq:BDC_encoder_LE}, i.e., \emph{message} and \emph{parity}. Note that coding schemes of \eqref{eq:BEC_decoder_LE} and \eqref{eq:BDC_encoder_LE} are based on the generator matrix. In addition, we can compare the duality of \emph{codeword elements of erasures} and \emph{codeword elements of normal cells} from \eqref{eq:BDC_encoder_LE} and \eqref{eq:BDC_encoder_LE_binning} which are coding schemes based on the parity check matrix.

In the BEC, the minimum distance $d_{\mathrm{min}}$ is defined by the \emph{parity check matrix} $H$, whereas the minimum distance $d_0$ of additive encoding for the BDC is defined by the \emph{generator matrix} $G_0$. The upper bound on the probability of decoding failure given $e$ erasures is dependent on the weight distribution of $\mathcal{C}$, whereas the upper bound on the probability of masking failure given $u$ defects is dependent on the weight distribution of $\mathcal{C}_0^{\perp}$.

If $A_w = B_w$ and $e = u$, it is clear that the upper bound on $P\left( D=0 \mid |\mathcal{E}|=e \right)$ is same as the upper bound on $P\left( M=0 \mid |\mathcal{U}|=u \right)$ by Theorem~\ref{thm:BEC_bound} and Theorem~\ref{thm:BDC_bound}. In particular, the following Theorem shows the equivalence of the failure probabilities (i.e., the probability of correction failure of erasures and the probability of masking failure of defects).

\begin{theorem} If $A_w = B_w$ and $\alpha = \beta$, then the probability of decoding failure of the BEC is same as the probability of masking failure of the BDC.\label{thm:BEC_BDC_failure}
\end{theorem}
\begin{IEEEproof}
Without loss of generality, we can assume that the all-zero codeword $\mathbf{0}$ has been transmitted through the BEC. If there is only one nonzero codeword such that $\Psi(\mathbf{c}_1) \subseteq \mathcal{E}$ where $\mathcal{E}$ indicates the location of $e$ erasures, the decoding success probability $P \left(D=1 \mid |\mathcal{E}|=e\right)=\frac{1}{2}$ since the decoder chooses between $\mathbf{0}$ and $\mathbf{c}_1$ randomly.

If there are two nonzero codewords $\mathbf{c}_1$ and $\mathbf{c}_2$ such that $\Psi(\mathbf{c}_1) \subseteq \mathcal{E}$ and $\Psi(\mathbf{c}_2) \subseteq \mathcal{E}$, it is clear that $\Psi(\mathbf{c}_3) \subseteq \mathcal{E}$ where $\mathbf{c}_3 = \mathbf{c}_1 + \mathbf{c}_2$. Since the decoder chooses one codeword from $\left\{\mathbf{0} , \mathbf{c}_1, \mathbf{c}_2, \mathbf{c}_3 \right\}$ randomly, $P \left(D=1 \mid |\mathcal{E}|=e \right)=\frac{1}{4}$. Similarly, if there are three codewords $\mathbf{c}_i$ such that $\Psi(\mathbf{c}_i) \subseteq \mathcal{E}$ for $i=1, 2, 3$ and $\mathbf{c}_1 + \mathbf{c}_2 \ne \mathbf{c}_3$, $P \left(D=1 \mid |\mathcal{E}|=e \right)=\frac{1}{8}$ since the decoder randomly chooses a codeword among $\left\{\mathbf{0} , \mathbf{c}_1, \mathbf{c}_2, \mathbf{c}_3, \mathbf{c}_1 + \mathbf{c}_2, \mathbf{c}_1 + \mathbf{c}_3, \mathbf{c}_2 + \mathbf{c}_3, \mathbf{c}_1 +\mathbf{c}_2 + \mathbf{c}_3 \right\}$. Generalizing this observation, we can claim that
\begin{equation} \label{eq:duality_BEC}
P \left(D=1 \mid |\mathcal{E}|=e \right)=\frac{1}{2^j}
\end{equation}
if $\Psi(\mathbf{c}_i) \subseteq \mathcal{E}$ for $i = 0,1\ldots,2^j-1$ and $\mathbf{c}_0 = \mathbf{0}$.

It is clear there are at least $j$ codewords $\mathbf{c}^{\perp} \in \mathcal{C}_0^{\perp}$ such that $G_0^T \mathbf{c}^{\perp} = \mathbf{0}$ and $\Psi(\mathbf{c}^{\perp}) \subseteq \mathcal{U}$ for $\rank\left( G_0^{\mathcal{U}} \right) = u - j$. From these $j$ codewords, we can list $2^j$ codewords $\mathbf{c}^{\perp} \in \mathcal{C}_0^{\perp}$ such that $G_0^T \mathbf{c}^{\perp} = \mathbf{0}$ and $\Psi(\mathbf{c}^{\perp}) \subseteq \mathcal{U}$. Since the last $j$ elements of the column vector $\mathbf{b}^{\mathcal{U}}$ in \eqref{eq:BDC_LE_sol_exist} should be zeros for masking success, we can claim that
\begin{equation} \label{eq:duality_BDC}
P\left(M=1 \mid |\mathcal{U}|=u \right) = \frac{1}{2^j}
\end{equation}
if $\Psi(\mathbf{c}_i^{\perp}) \subseteq \mathcal{U}$ for $i = 0,1\ldots,2^j-1$. It is assumed that the distribution of each element of $\mathbf{b}^{\mathcal{U}}$ is uniform since $P(S=0 \mid S \ne \lambda) = P(S=1 \mid S \ne \lambda) = \frac{1}{2}$.

If $\alpha = \beta$, the number of erasures $|\mathcal{E}|$ and the number of defects $|\mathcal{U}|$ follow an identical binomial distribution. If $A_w = B_w$, the codeword set $\mathcal{C}$ for the BEC and the dual codeword set $\mathcal{C}_0^{\perp}$ for the BDC are also identical. Thus, we can claim that $P(D = 1 \mid |\mathcal{E}|=e) = P(M = 1 \mid |\mathcal{U}|=u)$ by \eqref{eq:duality_BEC} and \eqref{eq:duality_BDC}.

Since
\begin{align}
P(D=1) &= \sum{P\left( |\mathcal{E}| = e \right)}{P(D = 1 \mid |\mathcal{E}|=e)}, \\
P(M=1) &= \sum{P\left( |\mathcal{U}| = u \right)}{P(U = 1 \mid |\mathcal{U}|=u)},
\end{align}
we can claim that $P(D=1) = P(M=1)$.
\end{IEEEproof}
In Section~\ref{subsection:BEC_BDC}, we show that numerical results confirm the duality between erasures and defects.

From channel properties, capacities, capacity achieving schemes, their upper bounds, and their failure probability, we have demonstrated the duality between erasures and defects.

\begin{table}[t]
\renewcommand{\arraystretch}{1.3}
\caption{Duality between BEC and BDC}
\label{tab:duality}
\centering
{\small
\hfill{}
\begin{tabular}{|c|c|c|}
\hline
                     & BEC & BDC  \\ \hline \hline
Channel property     & Ternary output $Y \in \{0, 1, \varepsilon \}$ & Ternary state $S \in \{0, 1, \lambda\}$  \\ \hline
Value  & Erasure $\varepsilon$ is neither ``0'' nor ``1'' & Defect is either ``0'' or ``1'' \\ \hline
Capacity             & $C_{\mathrm{BEC}} = 1 - \alpha$ \: \eqref{eq:BEC_capacity} & $C_{\mathrm{BDC}} = 1 - \beta$ \: \eqref{eq:BDC_capacity} \\ \hline
Channel information  & Locations & Locations and stuck-at values \\ \hline
Correcting / masking & Decoder corrects erasures & Encoder masks defects    \\ \hline
\multirow{2}{*}{Linear equation} &  $G^{\mathcal{V}} \widehat{\mathbf{m}} =\mathbf{y}^{\mathcal{V}}$\: \eqref{eq:BEC_decoder_LE} & $G_0^{\mathcal{U}} \mathbf{d} = \mathbf{b}^{\mathcal{U}}$ \:\eqref{eq:BDC_encoder_LE} \\
 & $\left(H^{\mathcal{E}} \right)^T \widehat{\mathbf{c}}^{\mathcal{E}} = \mathbf{q}$ \ \eqref{eq:BEC_decoder_LE_H} & $ \left(H_0^{\mathcal{W}} \right)^T \mathbf{c}^{\mathcal{W}} = \mathbf{m}'$ \ \eqref{eq:BDC_encoder_LE_binning}  \\ \hline
\multirow{2}{*}{Solution of linear equation} & $\widehat{\mathbf{m}}$ (estimate of message) or & $\mathbf{d}$ (parity) or \\
 & $\widehat{\mathbf{c}}^{\mathcal{E}}$ (estimate of erased bits) & $\mathbf{c}^{\mathcal{W}}$ (codeword elements of normal cells) \\ \hline
Type of linear equation & Overdetermined & Underdetermined \\ \hline
\multirow{2}{*}{Minimum distance} & $d_{\text{min}} = \min\{ \|\mathbf{c}\|: H^T \mathbf{c} = \mathbf{0}, \mathbf{c} \ne \mathbf{0} \}$ & $d_{0} = \min\{ \|\mathbf{c}\|: G_0^T \mathbf{c} = \mathbf{0}, \mathbf{c} \ne \mathbf{0} \}$ \\
 & If $e < d_{\text{min}}$, $e$ erasures are corrected. & If $u < d_0$, $u$ defects are masked. \\ \hline
Upper bound on& \multirow{2}{*}{Theorem~\ref{thm:BEC_bound}}  & \multirow{2}{*}{Theorem~\ref{thm:BDC_bound}} \\
probability of failure & & \\ \hline
Probability of failure & \multicolumn{2}{c|}{If $A_w = B_w$ and $\alpha = \beta$, then $P(D = 0) = P (M = 0)$ (Theorem~\ref{thm:BEC_BDC_failure}) } \\ \hline

\end{tabular}}
\hfill{}
\end{table}

\section{Binary Defect and Erasure Channel}\label{sec:BDEC}

\subsection{Binary Defect and Erasure Channel}

Considering the duality between erasures and defects, we now introduce the BDEC which has both erasures and defects. As shown in Fig.~\ref{fig:BDEC}, the probability of defects are defined by \eqref{eq:BDC_random_channel_model}, and normal cells behave as the BEC with parameter $\alpha$. The capacity of the BDEC was given by \eqref{eq:BDEC_capacity}.

In order to mask defects and correct erasures, the following two cases will be considered.

\begin{itemize}
  \item Case 1: Only the decoder has knowledge of both defects and erasures.
  \item Case 2: The encoder has only knowledge of defects and the decoder has only knowledge of erasures.
\end{itemize}

In case 1, the decoder can regard defects as erasures. Then, a fraction $\left(1 - \beta \right)\left(1 - \alpha \right)$ of bits are unerased. The coding scheme for the BEC can achieve the channel capacity of \eqref{eq:BDEC_capacity} since the BDEC is equivalent to the BEC with parameter $1 - \left(1 - \beta \right)\left(1 - \alpha \right)$.

In case 2, the encoder masks the defects by the additive encoding and the decoder corrects the erasures. The proposed coding scheme for the BDEC combines the encoding of the BDC and the decoding of the BEC.

\emph{Encoding:} A message $\mathbf{m} \in \{0, 1\}^k$ is encoded to a codeword $\mathbf{c}= G_1 \mathbf{m} + G_0 \mathbf{d}$. Note that $G_1$ is an $n \times k$ generator matrix and $G_0$ is an $n \times l$ generator matrix such that $n > k + l$. Two generator matrices are used to correct erasures and mask defects. First, $G_1$ encodes a message $\mathbf{m}$ into $G_1 \mathbf{m}$ for correcting erasures. Next, the defects will be masked by $G_0 \mathbf{d}$. The parity for masking defects $\mathbf{d}$ will be chosen by solving \eqref{eq:BDC_encoder_LE}. The encoding can be represented by
\begin{equation}\label{eq:BDEC_encoder}
\mathbf{c} = \widetilde{G} \begin{bmatrix} \mathbf{m} \\ \mathbf{d} \end{bmatrix}
\end{equation}
where $\widetilde{G} = \left[ G_1 \quad G_0 \right]$ is an $n \times (k + l)$ matrix. Note that $r = n - k - l$ is the number of parity bits for correcting erasures and $l$ is the number of parity bits for masking defects.

\emph{Decoding:} The decoding of BDEC can be done by solving the following linear equations.
\begin{equation}\label{eq:BDEC_decoder}
\widetilde{G}^{\mathcal{V}} \begin{bmatrix} \widehat{\mathbf{m}} \\ \widehat{\mathbf{d}} \end{bmatrix} = \mathbf{y}^{\mathcal{V}}
\end{equation}
where $\mathcal{V}=\left\{j_1,\cdots, j_v\right\}$ indicates the locations of $v$ unerased bits. We use the notation of $\mathbf{y}^{\mathcal{V}}=\left(y_{j_1}, \cdots, y_{j_v}\right)^T$ and $\widetilde{G}^{\mathcal{V}}=\left[ \widetilde{\mathbf{g}}_{j_1}^T, \cdots, \widetilde{\mathbf{g}}_{j_v}^T \right]^T$ where $\mathbf{g}_j$ is the $j$-th row of $\widetilde{G}$. By solving \eqref{eq:BDEC_decoder}, we can obtain the estimate of message $\mathbf{m}$ and the estimate of parity $\mathbf{d}$, i.e., $\widehat{\mathbf{m}}$ and $\widehat{\mathbf{d}}$. Note that \eqref{eq:BDEC_decoder} is equivalent to \eqref{eq:BEC_decoder_LE}.

Also, the decoding can be done by solving the following linear equations based on the parity check matrix $\widetilde{H}$ instead of \eqref{eq:BDEC_decoder}.
\begin{equation}
\left(\widetilde{H}^{\mathcal{E}} \right)^T \widehat{\mathbf{c}}^{\mathcal{E}} = \mathbf{q}'
\end{equation}
where $\mathbf{q}' = \left(\widetilde{H}^{\mathcal{V}} \right)^T \mathbf{y}^{\mathcal{V}}$.

The weight distribution of the coding scheme is defined as a pair of sets $\left(A_{1,w}, B_{0,w} \right)$. $A_{1, w}$ is the weight distribution of the $[n, k+l]$ linear block code with the generator matrix $\widetilde{G} = \left[G_1 \quad G_0 \right]$ and the parity check matrix $\widetilde{H}$. Also, $B_{0, w}$ is the weight distribution of the $[n, k+r]$ linear block code with parity check matrix $G_0$ \cite{Heegard1983}. Thus, \eqref{eq:Aw_UB} will be modified into
\begin{align}
A_{1,w} & \le \frac{n}{2^{n - (k+l)}} \binom{n}{w}, \label{eq:BDEC_A1w_UB}\\
B_{0,w} & \le \frac{n}{2^{n - (k+r)}} \binom{n}{w}. \label{eq:BDEC_B0w_UB}
\end{align}

Also, a pair of minimum distances $(d_0 , d_1)$ are defined, where $d_0$ represents the minimum distance for masking defects and $d_1$ is the minimum distance for correcting erasures in the BDEC. $d_0$ is same as \eqref{eq:BDC_dmin} and $d_1$ is given by
\begin{align}
d_1 &= \underset{
\substack{
\mathbf{m}  \ne \mathbf{0} \\
\widetilde{H}^T \mathbf{c}= \mathbf{0}
}}
{\text{min }} \|\mathbf{c}\| \label{eq:BDEC_d1}
\end{align}
Note that $d_1$ is greater than or equal to the minimum distance of the $[n,k+l]$ linear block code with parity check matrix $\widetilde{H}$, while $d_0$ is the minimum distance of the $[n,k+r]$ linear block code with the parity check matrix $G_0$ \cite{Heegard1983}. 

In case 2, the encoder solves the linear equations of \eqref{eq:BDC_encoder_LE} in order to determine the parity $\mathbf{d}$ for masking defects. Also, the decoder solves the linear equations of \eqref{eq:BDEC_decoder} to estimate $\mathbf{m}$. Thus, it is clear that this coding scheme is a combination of the coding scheme for the BEC and the coding scheme for the BDC. We will now prove that this proposed coding scheme is a capacity achieving scheme.
%
%
%

\begin{theorem} The proposed coding scheme achieves the capacity of the BDEC. The encoding and the decoding are represented by \eqref{eq:BDEC_encoder} and \eqref{eq:BDEC_decoder}, respectively. \label{thm:BDEC_CAS}
\end{theorem}

\begin{IEEEproof}
We can see that
\begin{align}
P\left( \widehat{\mathbf{m}} \ne \mathbf{m} \right) & = P\left(M=0, D=1\right) + P\left(M=0, D=0\right) + P\left(M=1, D=0\right) \\
    & = P\left(M=0\right) + P\left(M=1, D=0\right). \label{eq:BDEC_P_failure}
\end{align}
First, we will derive the upper bound on $P\left(M=0\right)$, which is similar to Theorem~\ref{thm:BDC}. The only difference is that $B_{0, w}$ of \eqref{eq:BDEC_B0w_UB} should be used instead of $B_w$. Thus, \eqref{eq:BDC_CAS_capacity} will be changed into
\begin{equation}
P(M=0) \le n^2 (\beta + \epsilon) 2^{n \left\{ \frac{k+r}{n} - (1 - \beta) + \epsilon \right\}} + \epsilon'. \label{eq:BDEC_CAS_UB_M}
\end{equation}

Next, the upper bound on $P\left(M = 1, D = 0\right)$ will be derived.
\begin{align}
& P\left(M = 1, D = 0\right) \\
    & = P \left(M = 1, D = 0, U \le n(\beta + \epsilon), E \le n\left\{(1 - \beta) \alpha + \epsilon \right\} \right) + \epsilon' \\
    & = \sum_{u=0}^{n(\beta + \epsilon)}{\sum_{e=0}^{n((1 - \beta) \alpha + \epsilon)}{P \left(M = 1, D = 0, |\mathcal{U}|=u, |\mathcal{E}|=e \right)}} + \epsilon' \\
    & = \sum_{u=0}^{n(\beta + \epsilon)}{\sum_{e=0}^{n((1 - \beta) \alpha + \epsilon)}{P(|\mathcal{U}|=u) P(M = 1 \mid |\mathcal{U}|=u) P(|\mathcal{E}|=e \mid M = 1, |\mathcal{U}|=u)}} \nonumber \\
    & \qquad \qquad \qquad \quad \ \cdot P(D = 0 \mid M = 1, |\mathcal{U}|=u, |\mathcal{E}|=e) + \epsilon' \label{eq:BDEC_CAS_chain_rule} \\
    & \le \sum_{u=0}^{n(\beta + \epsilon)}{\sum_{e=0}^{n((1 - \beta) \alpha + \epsilon)}{P(D = 0 \mid M = 1, |\mathcal{U}|=u, |\mathcal{E}|=e) }} + \epsilon' \\
    & \le \sum_{u=0}^{n(\beta + \epsilon)}{\sum_{e=0}^{n((1 - \beta) \alpha + \epsilon)}{ \frac{ \sum_{w = d_1}^{e}{A_{1,w} \binom{n-u-w}{e-w}} }{\binom{n-u}{e}} }} + \epsilon' \label{eq:BDEC_CAS_UB}\\
    & \le \frac{n}{2^{n - (k+l)}} \sum_{u=0}^{n(\beta + \epsilon)}{\sum_{e=0}^{n((1 - \beta) \alpha + \epsilon)}{ \frac{ \sum_{w = d_1}^{e}{ \binom{n}{w} \binom{n-u-w}{e-w}} }{\binom{n-u}{e}} }} + \epsilon' \label{eq:BDEC_CAS_UB_given_e} \\
    & \le \frac{n}{2^{n - (k+l)}} \sum_{u=0}^{n(\beta + \epsilon)}{\sum_{e=0}^{n((1 - \beta) \alpha + \epsilon)}{ \sum_{w = d_1}^{e}{\binom{e}{w}} \frac{\binom{n}{w}}{\binom{n-u}{w}} }} + \epsilon' \label{eq:BDEC_CAS_binom} \\
    & \le \frac{n}{2^{n - (k+l)}} \sum_{u=0}^{n(\beta + \epsilon)}{\sum_{e=0}^{n((1 - \beta) \alpha + \epsilon)}{ \sum_{w = d_1}^{e}{\binom{e}{w}}}} + \epsilon' \label{eq:BDEC_CAS_lemma_binom} \\
    & \le \frac{n}{2^{n - (k+l)}} \sum_{u=0}^{n(\beta + \epsilon)}{\sum_{e=0}^{n((1 - \beta) \alpha + \epsilon)}{ 2^e }} + \epsilon' \\
    & \le \frac{n}{2^{n - (k+l)}} \sum_{u=0}^{n(\beta + \epsilon)}{\sum_{e=0}^{n((1 - \beta) \alpha + \epsilon)}{ 2^{n((1 - \beta) \alpha + \epsilon)} }} + \epsilon' \\
    & \le n^3 \left\{ (\beta + \epsilon) + \frac{1}{n} \right\} \left\{\left( \left( 1 - \beta \right) \alpha + \epsilon \right) + \frac{1}{n}\right\} { 2^{n \left\{\frac{k+l}{n} - 1 + (1 - \beta) \alpha + \epsilon \right\}} } + \epsilon' \label{eq:BDEC_CAS_UB_D}
\end{align}
where we assume that $n(\beta + \epsilon)$ and $n((1 - \beta) \alpha + \epsilon)$ are integers without loss of generality. \eqref{eq:BDEC_CAS_chain_rule} follows from the chain rule. \eqref{eq:BDEC_CAS_UB} follows from the modification of \eqref{eq:BEC_UB} where all the defects are successfully masked and we do not need to consider the defects. Also, $A_{1, w}$ of \eqref{eq:BDEC_A1w_UB} has been used instead of $A_w$. \eqref{eq:BDEC_CAS_binom} follows from ${\binom{n-u-w}{e-w}}/{\binom{n-u}{e}} = {\binom{e}{w}}/{\binom{n-u}{w}}$. In addition, \eqref{eq:BDEC_CAS_lemma_binom} follows from ${\binom{n}{w}}/{\binom{n-u}{w}}\le 1$ for $0 \le u \le n$.

By \eqref{eq:BDEC_CAS_UB_M} and \eqref{eq:BDEC_CAS_UB_D}, we can see that the upper bound on $P\left( \widehat{\mathbf{m}} \ne \mathbf{m} \right)$ goes to zero if $n$ is sufficiently large and the following two conditions hold.
\begin{align}
\frac{k+r}{n} & < \left(1 - \beta \right) - \epsilon \label{eq:BDEC_CAS_C1} \\
\frac{k+l}{n} & < 1 - \left(1 - \beta \right) \alpha - \epsilon  \label{eq:BDEC_CAS_C2}
\end{align}
From the sum of \eqref{eq:BDEC_CAS_C1} and \eqref{eq:BDEC_CAS_C2},
\begin{equation}
R = \frac{k}{n} < (1 - \beta)(1 - \alpha) - 2 \epsilon = C_{\mathrm{BDEC}} - 2\epsilon.
\end{equation}
Thus, the proposed coding scheme achieves the channel capacity of BDEC.
\end{IEEEproof}

\subsection{Redundancy Allocation of BDEC}

The proposed coding scheme for the BDEC requires two generator matrices, namely $G_0$ for masking defects and $G_1$ for correcting erasures, which results in two parts of redundancy. Since the number of parity bits for masking defects and for correcting erasures are $l$ and $r$ respectively, the total redundancy is $l+r = n-k$ and the code rate is $R = k/n$.

The fact that the redundancy can be divided into two parts leads to the problem of redundancy allocation. The objective is to find an optimal redundancy allocation between $l$ and $r$ in order to minimize $P\left( \widehat{\mathbf{m}} \ne \mathbf{m} \right)$. The problem of redundancy allocation can be formulated as follows \cite{Kim2013:plbc}.
\begin{equation}\label{eq:BDEC_opt_problem}
\begin{aligned}
( \widehat{l}, \widehat{r} )  = \; & \underset{(l, r)}{\text{argmin}} & &  P\left( \widehat{\mathbf{m}} \ne \mathbf{m} \right) \\
& \text{subject to} & &  l+r = n-k  \\
& & & 0 \le l \le n-k  \\
& & & 0 \le r \le n-k
\end{aligned}
\end{equation}

Not surprisingly, the optimal redundancy allocation depends on the BDEC parameters $\alpha$ and $\beta$. For the BDC (i.e., $\alpha$ = 0), we should allot all redundancy to masking defects and the optimal redundancy allocation will be $(l^{*}, r^{*})=(n-k, 0)$. Meanwhile, the optimal redundancy allocation for the BEC (i.e., $\beta$ = 0) will be $(l^{*}, r^{*})=(0, n-k)$, which is same as the result of \cite{Kim2013:plbc}.

When the BDEC has both defects and erasures (i.e., $\alpha \ne 0$ and $\beta \ne 0$), it is not straightforward to obtain the optimal redundancy allocation $(l^{*}, r^{*})$. Without an expression for $P\left( \widehat{\mathbf{m}} \ne \mathbf{m} \right)$ as a function of $\left(l, r\right)$, this optimization problem cannot be solved. Unfortunately, it is difficult to obtain the exact mathematical expression for $P\left( \widehat{\mathbf{m}} \ne \mathbf{m} \right)$.

Alternatively, we can obtain $(l^*, r^*)$ via Monte-Carlo simulations. However, to find $(l^{*}, r^{*})$ by simulations requires significant computations, especially for a low $P\left( \widehat{\mathbf{m}} \ne \mathbf{m} \right)$. Thus, we will consider an estimate $(\widehat{l}, \widehat{r})$ which minimizes the upper bound on $P\left( \widehat{\mathbf{m}} \ne \mathbf{m} \right)$ instead of $P\left( \widehat{\mathbf{m}} \ne \mathbf{m} \right)$.

For sufficiently large $n$, the upper bound on $P\left( \widehat{\mathbf{m}} \ne \mathbf{m} \right)$ was already derived in Theorem~\ref{thm:BDEC_CAS} since the upper bound is the sum of \eqref{eq:BDEC_CAS_UB_M} and \eqref{eq:BDEC_CAS_UB_D}. From \eqref{eq:BDEC_CAS_C1} and \eqref{eq:BDEC_CAS_C2} in Theorem~\ref{thm:BDEC_CAS}, the required redundancy $\left(l, r\right)$ for achieving the capacity can be given by
\begin{align}
l & > n \left( \beta + \epsilon \right)  \label{eq:BDEC_CAS_C1_modified}, \\
r & > n \left\{ \left(1 - \beta \right) \alpha + \epsilon \right\}.  \label{eq:BDEC_CAS_C2_modified}
\end{align}
However, these asymptotic results are not useful to choose the redundancy allocation of $\left(l, r\right)$ for a finite length code. Thus, we will derive the upper bound on $P\left( \widehat{\mathbf{m}} \ne \mathbf{m} \right)$ for a finite $n$.

We assume that the weight distributions $A_{1,w}$ and $B_{0,w}$ can be approximated by the binomial distribution as follows.
\begin{align}
A_{1,w} & \cong 2^{-r} \binom{n}{w} \label{eq:BDEC_A1w}\\
B_{0,w} & \cong 2^{-l} \binom{n}{w} \label{eq:BDEC_B0w}
\end{align}
which hold for random codes. In addition, the weight distribution of BCH codes can be approximated by the above binomial distribution \cite{Macwilliams1977theory}. By using \eqref{eq:BDEC_A1w} and \eqref{eq:BDEC_B0w} instead of \eqref{eq:BDEC_A1w_UB} and \eqref{eq:BDEC_B0w_UB}, the upper bound on $P\left( \widehat{\mathbf{m}} \ne \mathbf{m} \right)$ for a finite $n$ will be derived in the following Theorem.

\begin{theorem} \label{thm:RA_UB}For a finite $n$, the upper bound on $P\left( \widehat{\mathbf{m}} \ne \mathbf{m} \right)$ of the BDEC is given by
\begin{equation} \label{eq:RA_UB}
P\left( \widehat{\mathbf{m}} \ne \mathbf{m} \right) \le 2^{-l} \left(1 + \beta \right)^n + 2^{-r} \left\{ 1 + \alpha \left(1 - \beta \right) \right\}^n.
\end{equation}
\end{theorem}

\begin{IEEEproof}
The proof for the finite $n$ is similar to the proof of Theorem~\ref{thm:BDEC_CAS}. 
First, the upper bound on $P(M=0)$ is given by
\begin{align}
P\left(M=0\right) & = \sum_{u=0}^{n}{P(|\mathcal{U}|=u) P\left(M=0 \mid |\mathcal{U}|=u \right)} \label{eq:RA_chain_rule1} \\
    & \le \sum_{u = d_0}^{n}{\binom{n}{u} \beta^u \left(1 - \beta \right)^{n-u}  \frac{\sum_{w=d_0}^{u}{B_{0, w} \binom{n-w}{u-w}}}{\binom{n}{u}} } \label{eq:RA_B_1} \\
    & = 2^{-l} \sum_{u = d_0}^{n}{\beta^u \left(1 - \beta \right)^{n-u} \sum_{w=d_0}^{u}{\binom{n}{w} \binom{n-w}{u-w}}} \label{eq:RA_B_2} \\
    & = 2^{-l} \sum_{u = d_0}^{n}{\beta^u \left(1 - \beta \right)^{n-u} \sum_{w=d_0}^{u}{\binom{u}{w} \binom{n}{u}}} \label{eq:RA_binom_1} \\
    & \le 2^{-l} \sum_{u = 0}^{n}{\binom{n}{u} \beta^u \left(1 - \beta \right)^{n-u} \sum_{w = 0}^{u}{\binom{u}{w} }} \\
    & = 2^{-l} \sum_{u = 0}^{n}{\binom{n}{u} \left(2\beta\right)^u \left(1 - \beta \right)^{n-u}} \\
    & = 2^{-l} \left(1 + \beta \right)^n \label{eq:RA_M}
\end{align}
where \eqref{eq:RA_B_1} follows from \eqref{eq:BDC_UB} in Lemma~\ref{lemma:BDC_UB} and \eqref{eq:RA_B_2} comes from \eqref{eq:BDEC_B0w}. Also, \eqref{eq:RA_binom_1} follows from $\binom{n}{w} \binom{n-w}{u-w} = \binom{u}{w} \binom{n}{u}$.

Next, the upper bound on $P\left(M=1, D=0\right)$ is given by
\begin{align}
& P\left(M=1, D=0\right) \nonumber \\
& = \sum_{u=0}^{n}{ \sum_{e=0}^{n-u}{ P\left( M=1, D=0, |\mathcal{U}|=u, |\mathcal{E}|=e  \right)}} \\
& \le \sum_{u=0}^{n}{ \sum_{e=0}^{n-u}{ P\left( |\mathcal{U}|=u \right) P\left( |\mathcal{E}|=e \mid |\mathcal{U}|=u \right) P\left( D=0 \mid M=1, |\mathcal{U}|=u, |\mathcal{E}|=e  \right)}} \label{eq:RA_chain_rule2}\\
& \le \sum_{u=0}^{n}{  \binom{n}{u} \beta^u \left( 1-\beta \right)^{n-u} \sum_{e=0}^{n-u}{ \binom{n-u}{e} \alpha^e \left( 1-\alpha \right)^{n-u-e} \frac{ \sum_{w=d_1}^{e}{A_{1, w} \binom{n-u-w}{e-w}} }{\binom{n-u}{e}}}} \label{eq:RA_A_1} \\
& = 2^{-r} \sum_{u=0}^{n}{  \binom{n}{u} \beta^u \left( 1-\beta \right)^{n-u} \sum_{e=0}^{n-u}{ \binom{n-u}{e} \alpha^e \left( 1-\alpha \right)^{n-u-e} \frac{ \sum_{w=d_1}^{e}{\binom{n}{w} \binom{n-u-w}{e-w}} }{\binom{n-u}{e}}}} \label{eq:RA_A_2} \\
& = 2^{-r} \sum_{u=0}^{n}{  \binom{n}{u} \beta^u \left( 1-\beta \right)^{n-u} \sum_{e=0}^{n-u}{ \binom{n-u}{e} \alpha^e \left( 1-\alpha \right)^{n-u-e}  \sum_{w=d_1}^{e}{\binom{e}{w}\frac{\binom{n}{w}} {\binom{n-u}{w}}}}} \label{eq:RA_binom} \\
& \le 2^{-r} \sum_{u=0}^{n}{  \binom{n}{u} \beta^u \left( 1-\beta \right)^{n-u} \sum_{e=0}^{n-u}{ \binom{n-u}{e} \alpha^e \left( 1-\alpha \right)^{n-u-e}  \sum_{w=d_1}^{e}{\binom{e}{w}} }}\\
& \le 2^{-r} \sum_{u=0}^{n}{  \binom{n}{u} \beta^u \left( 1-\beta \right)^{n-u} \sum_{e=0}^{n-u}{ \binom{n-u}{e} (2\alpha)^e \left( 1-\alpha \right)^{n-u-e} }}\\
& \le 2^{-r} \sum_{u=0}^{n}{  \binom{n}{u} \beta^u \left( 1-\beta \right)^{n-u} \left( 1+ \alpha\right)^{n-u}}\\
& = 2^{-r} \sum_{u=0}^{n}{  \binom{n}{u} \beta^u \left\{ \left( 1-\beta \right)\left( 1+ \alpha\right) \right\}^{n-u} }\\
& = 2^{-r} \left\{1 + \alpha \left( 1- \beta\right) \right\}^{n} \label{eq:RA_D}
\end{align}
where \eqref{eq:RA_chain_rule2} follows from the chain rule and $P\left( M=1 \mid |\mathcal{E}|=e, |\mathcal{U}|=u \right) \le 1$. \eqref{eq:RA_A_1} follows from \eqref{eq:BEC_UB} in Lemma~\ref{lemma:BEC_UB} and \eqref{eq:RA_A_2} comes from \eqref{eq:BDEC_A1w}. In addition, \eqref{eq:RA_binom} is similar to \eqref{eq:BDEC_CAS_binom}.

Finally, \eqref{eq:RA_UB} is obtained from \eqref{eq:BDEC_P_failure}, \eqref{eq:RA_M}, and \eqref{eq:RA_D}.
\end{IEEEproof}

From Theorem~\ref{thm:RA_UB}, the upper bounds on $P\left( \widehat{\mathbf{m}} \ne \mathbf{m} \right)$ of the BEC and the BDC for a finite $n$ can be derived as follows.

\begin{corollary} \label{cor:BEC}For a finite $n$, the upper bound on $P\left( \widehat{\mathbf{m}} \ne \mathbf{m} \right)$ of the BEC is given by
\begin{align}
P\left( \widehat{\mathbf{m}} \ne \mathbf{m} \right) & \le 2^{-r} \left( 1 + \alpha \right)^n \label{eq:BEC_UB_finite}, \\
\log_2{P\left( \widehat{\mathbf{m}} \ne \mathbf{m} \right)} & \le n \left\{R - 1 + \log_2{(1+\alpha)} \right\}. \label{eq:BEC_UB_finite_log}
\end{align}
\end{corollary}
\begin{IEEEproof}
It is clear that $P(M=0)=0$ and $\beta = 0$ for the BEC. By \eqref{eq:BDEC_P_failure} and \eqref{eq:RA_D}, the upper bound on $P\left( \widehat{\mathbf{m}} \ne \mathbf{m} \right)$ of the BEC is given by
\begin{equation*}
P\left( \widehat{\mathbf{m}} \ne \mathbf{m} \right) = P\left(M=1, D=0\right) \le 2^{-r} \left( 1 + \alpha \right)^n.
\end{equation*}
Also, \eqref{eq:BEC_UB_finite_log} can be obtained by taking the logarithm.
\end{IEEEproof}

\begin{corollary} For a finite $n$, the upper bound on $P\left( \widehat{\mathbf{m}} \ne \mathbf{m} \right)$ of the BDC is given by
\begin{align}
P\left( \widehat{\mathbf{m}} \ne \mathbf{m} \right) & \le 2^{-l} \left( 1 + \beta \right)^n \label{eq:BDC_UB_finite} \\
\log_2{P\left( \widehat{\mathbf{m}} \ne \mathbf{m} \right)} & \le n \left\{R - 1 + \log_2{(1+\beta)} \right\}. \label{eq:BDC_UB_finite_log}.
\end{align}
\end{corollary}
\begin{IEEEproof}
It is clear that $P(M=1, D=0)=0$ and $\alpha = 0$ for the BDC. By \eqref{eq:BDEC_P_failure} and \eqref{eq:RA_M}, the upper bound on $P\left( \widehat{\mathbf{m}} \ne \mathbf{m} \right)$ of the BDC is given by
\begin{equation*}
P\left( \widehat{\mathbf{m}} \ne \mathbf{m} \right) = P\left(M=0\right) \le 2^{-l} \left( 1 + \beta \right)^n.
\end{equation*}
Also, \eqref{eq:BDC_UB_finite_log} can be obtained by taking the logarithm.
\end{IEEEproof}

Since $( \widehat{l}, \widehat{r} )$ minimizes the upper bound on $P\left( \widehat{\mathbf{m}} \ne \mathbf{m} \right)$, the optimization problem in \eqref{eq:BDEC_opt_problem} is given by
\begin{equation}\label{eq:BDEC_opt_problem_estimate}
\begin{aligned}
( \widehat{l}, \widehat{r} )  = \; & \underset{(l, r)}{\text{argmin}} & &  2^{-l} \left(1 + \beta \right)^n + 2^{-r} \left\{ 1 + \alpha \left(1 - \beta \right) \right\}^n \\
& \text{subject to} & &  l+r = n-k  \\
& & & 0 \le l \le n-k  \\
& & & 0 \le r \le n-k
\end{aligned}
\end{equation}
where the objective function is the upper bound on $P\left( \widehat{\mathbf{m}} \ne \mathbf{m} \right)$ for a finite $n$. This objective function is intuitively reasonable since $\beta$ is the probability of defects and $\alpha(1-\beta)$ is the probability of erasures. If $\beta \ge \alpha(1-\beta)$, we have to allot more redundancy for masking defects, i.e., $l \ge r$. Otherwise, we should allot more redundancy for correcting erasures. For $\alpha = 0$ or $\beta = 0$, we do not need to consider the above optimization problem since the solution of the BEC or the BDC is straightforward.

If the codeword length $n$, the information length $k$ and the channel parameters such as $\alpha$ and $\beta$ are given, the solution $( \widehat{l}, \widehat{r} )$ of the above optimization problem can be readily obtained. For example, we will consider $\left[ n = 1023, k=923, l \right]$ PBCH codes. All possible redundancy allocation candidates of PBCH codes are presented in Table~\ref{tab:plbc}. Since there are only 11 redundancy allocation candidates in Table~\ref{tab:plbc}, we can readily obtain the $( \widehat{l}, \widehat{r} )$ that minimizes the objective function of \eqref{eq:BDEC_opt_problem_estimate}.

\begin{table}[t]
\renewcommand{\arraystretch}{1.3}
\caption{All Possible Redundancy Allocation Candidates of $\left[ n = 1023, k=923, l \right]$ PBCH Codes}
\label{tab:plbc}
\centering
{\small
\begin{tabular}{|c|c|c|c|c|c|}
\hline
Code & {$l$} & {$r$} & {$d_0$} & {$d_1$} & Notes   \\ \hline \hline
0 & 0 & 100 & 0 & 21 & Only correcting erasures \\ \hline
1 & 10 & 90 & 3 & 19 &\\ \hline
2 & 20 & 80 & 5 & 17 &\\ \hline
3 & 30 & 70 & 7 & 15 &\\ \hline
4 & 40 & 60 & 9 & 13 &\\ \hline
5 & 50 & 50 & 11 & 11 & \\ \hline
6 & 60 & 40 & 13 & 9 &\\ \hline
7 & 70 & 30 & 15 & 7 &\\ \hline
8 & 80 & 20 & 17 & 5 &\\ \hline
9 & 90 & 10 & 19 & 3 &\\ \hline
10& 100 & 0 & 21 & 0 & Only masking defects\\ \hline
\end{tabular}}
\end{table}

In addition, the objective function is \emph{convex} if we assume that $l$ and $r$ are real values. Since the optimization problem is convex, we can derive the solution $(\widetilde{l}, \widetilde{r})$ of \eqref{eq:BDEC_opt_problem_estimate} by \emph{Karush-Kuhn-Tucker} (KKT) conditions.
\begin{numcases}{(\widetilde{l}, \widetilde{r})=}
(0, n-k), & if $2^{-l} \left(1 + \beta \right)^n \le 2^{-r} \left\{ 1 + \alpha \left(1 - \beta \right) \right\}^n$; \label{eq:BDEC_opt_sol_con1}
\\
(n-k, 0), & if $2^{-l} \left(1 + \beta \right)^n \ge 2^{-r} \left\{ 1 + \alpha \left(1 - \beta \right) \right\}^n$; \label{eq:BDEC_opt_sol_con2}
\\
\left( \check{l}, \check{r}  \right),  & \text{otherwise} \label{eq:BDEC_opt_sol_con3}
\end{numcases}
where $\left( \check{l}, \check{r} \right)$ is given by
\begin{align}
\check{l} &= \frac{1}{2} \left\{ n \left( 1 + \log_2{\left(\frac{1+\beta}{1 + \alpha (1 - \beta)}\right)} \right) - k \right\}, \label{eq:BDEC_opt_sol_1} \\
\check{r} &= \frac{1}{2} \left\{ n \left( 1 - \log_2{\left(\frac{1+\beta}{1 + \alpha (1 - \beta)}\right)} \right) - k \right\}. \label{eq:BDEC_opt_sol_2}
\end{align}
The details of derivation are given in Appendix. \eqref{eq:BDEC_opt_sol_con1} and \eqref{eq:BDEC_opt_sol_con2} are easy to see. Also, \eqref{eq:BDEC_opt_sol_1} and \eqref{eq:BDEC_opt_sol_2} are intuitively reasonable since $\check{l} \ge \check{r}$ for $\beta \ge \alpha(1 - \beta)$. If $\beta < \alpha(1 - \beta)$, $\check{l} < \check{r}$.

In Section~\ref{subsection:redundancy_allocation}, the numerical results show that $(\widehat{l}, \widehat{r})$ and $(\widetilde{l}, \widetilde{r})$ match $(l^*, r^*)$ very well.

\section{Numerical Results}\label{sec:numerical_results}

\subsection{BEC and BDC} \label{subsection:BEC_BDC}

The numerical results for the BEC and the BDC will be presented. For the BEC, the \emph{generator matrices} of BCH codes are used for $G$ of \eqref{eq:BEC_decoder_LE}. For the BDC, the PBCH codes are used, so the \emph{parity check matrices} of BCH codes are used for $G_0$ of \eqref{eq:BDC_encoder_LE} \cite{Heegard1983}. Thus, $A_w$ for the BEC and $B_w$ for the BDC are same.

\begin{figure}[!t]
   \centering
   \includegraphics[width=0.60\textwidth]{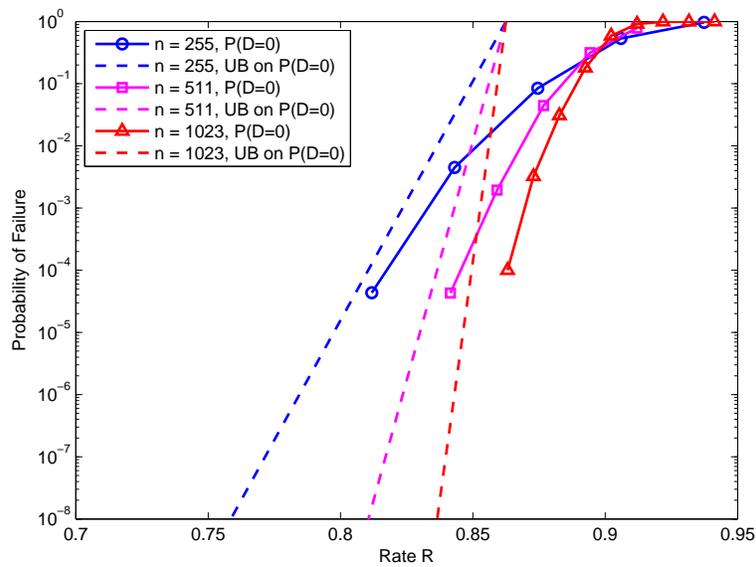}
   \caption{Probability of decoding failure, i.e., $P(D=0)$ for the BEC with $\alpha=0.1$.}
   \label{fig:BEC_numerical}
\end{figure}

\begin{figure}[!t]
   \centering
   \includegraphics[width=0.60\textwidth]{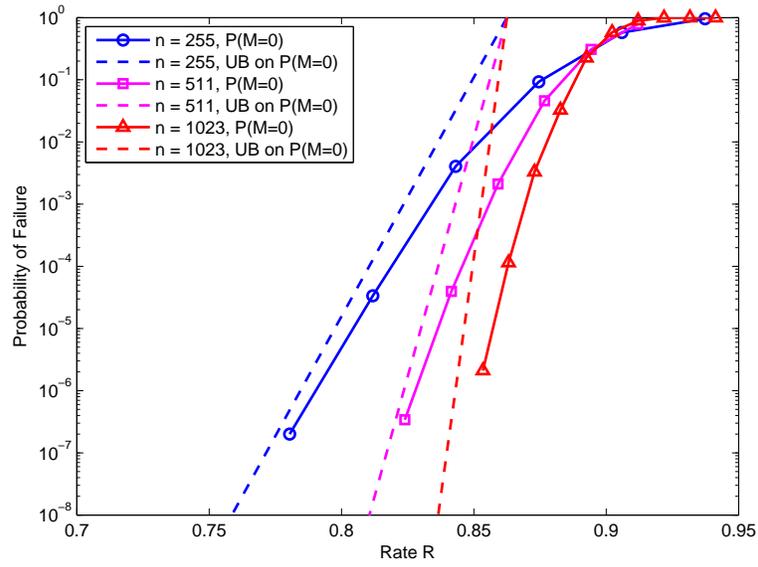}
   \caption{Probability of masking failure, i.e., $P(M=0)$ for the BDC with $\beta=0.1$.}
   \label{fig:BDC_numerical}
\end{figure}

Fig.~\ref{fig:BEC_numerical} shows the probability of decoding failure (i.e., $P(D=0)$) and its upper bound. Also, Fig.~\ref{fig:BDC_numerical} shows the probability of masking failure (i.e., $P(M=0)$) and its upper bound. The upper bounds are given by \eqref{eq:BEC_UB_finite} and \eqref{eq:BDC_UB_finite}. Since $\alpha = \beta = 0.1$, the upper bound for the BEC is same as the upper bound for the BDC.

Note that the slope of the upper bound on $ \log_2{P\left( \widehat{\mathbf{m}} \ne \mathbf{m} \right)}$ is $n$ as shown in Fig.~\ref{fig:BEC_numerical} and Fig.~\ref{fig:BDC_numerical}, which can be explained by \eqref{eq:BEC_UB_finite_log} and \eqref{eq:BDC_UB_finite_log}. Also, the x-axis intercepts (i.e., $R$ for $ \log_2{P\left( \widehat{\mathbf{m}} \ne \mathbf{m} \right)} = 0$) are $1 - \log_2{\left(1 + \alpha\right)}$ and $1 - \log_2{\left(1 + \beta\right)}$ for each channel.

\begin{figure}[!t]
   \centering
   \includegraphics[width=0.60\textwidth]{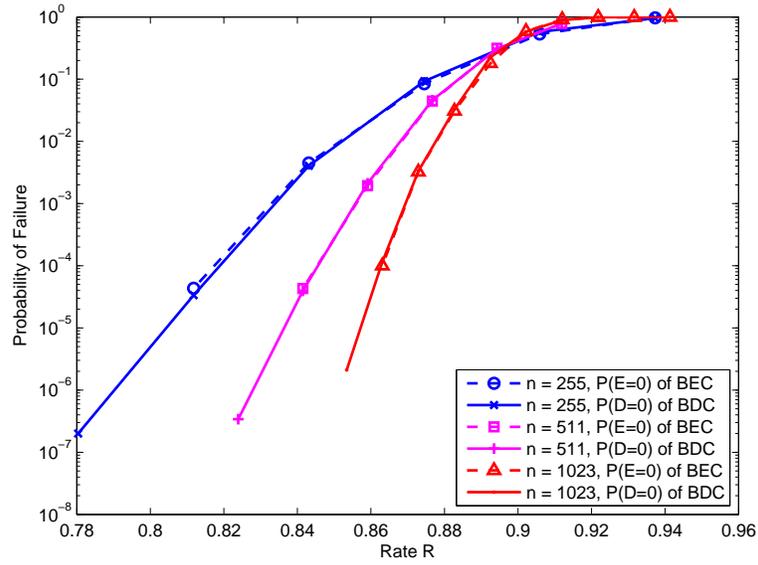}
   \caption{Probability of failure, i.e., $P(D=0)$ for the BEC with $\alpha=0.1$ and $P(M=0)$ for the BDC with $\beta=0.1$.}
   \label{fig:BEC_BDC_numerical}
\end{figure}

Fig.~\ref{fig:BEC_BDC_numerical} compares the probability of decoding failure of the BEC and the probability of masking failure of the BDC. By Fig.~\ref{fig:BEC_BDC_numerical}, we can see that $P(D=0) = P(M=0)$ if $A_w = B_w$ and $\alpha = \beta$, which confirms the duality in Theorem~\ref{thm:BEC_BDC_failure}.


\subsection{Redundancy Allocation for BDEC} \label{subsection:redundancy_allocation}

\begin{table}[t]
\renewcommand{\arraystretch}{1.3}
\caption{BDEC with the Same $C_{\textrm{BDEC}}=0.95$}
\label{tab:channel}
\centering
{\small
\hfill{}
\begin{tabular}{|c|c|c|c|}
\hline
Channel & {$\alpha$} & {$\beta$} & {Notes} \\ \hline \hline
1       & 0.0500 & 0 & BEC \\ \hline
2       & 0.0404 & 0.0100 &  \\ \hline
3       & 0.0306 & 0.0200 &  \\ \hline
4       & 0.0253 & 0.0253 &  \\ \hline
5       & 0.0200 & 0.0306 &  \\ \hline
6       & 0.0100 & 0.0404 &  \\ \hline
7       & 0 & 0.0500 & BDC \\  \hline
\end{tabular}}
\hfill{}
\end{table}

In order to discuss the redundancy allocation for BDEC, we will consider multiple BDECs in Table~\ref{tab:channel} whose capacities are $C_{\textrm{BDEC}}=0.95$. For these channels, we apply $\left[ n = 1023, k=923, l \right]$ PBCH codes whose all possible redundancy allocation candidates are presented in Table~\ref{tab:plbc}.

\begin{figure}[!t]
   \centering
   \includegraphics[width=0.60\textwidth]{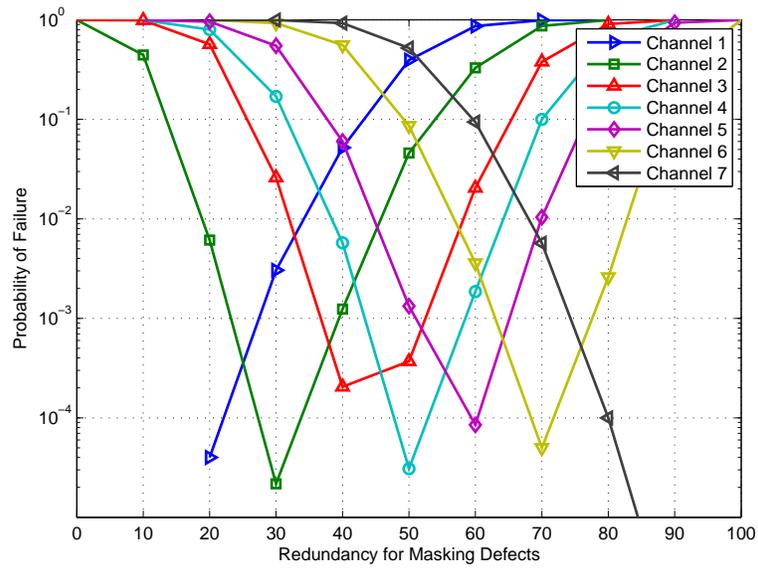}
   \caption{Probability of failure for the channels in Table~\ref{tab:channel}.}
   \label{fig:redundancy_allocation_simulation}
\end{figure}

Fig.~\ref{fig:redundancy_allocation_simulation} shows the simulation results for the channels of Table~\ref{tab:plbc}. The simulation results of channel 1 (BEC) and channel 7 (BDC) are incomplete due to their impractical computational complexities. However, it should be obvious that the optimal redundancy allocation for channel 1 (BEC) will be $\left(l^*, r^* \right) = (0, 100)$. The more defects a channel has, the larger $l$ is expected to be for the optimal redundancy allocation. Eventually, the optimal redundancy allocation for channel 7 (BDC) will be $\left(l^*, r^* \right) = (100, 0)$. The optimal $l^*$ for all channels of Table~\ref{tab:channel} can be obtained from Fig.~\ref{fig:redundancy_allocation_simulation}, which are presented in the second column of Table~\ref{tab:redundancy}. The optimal $r^*$ can be obtained by $r^* = n - k - l^*$ \cite{Kim2013:plbc}.

To find the optimal redundancy allocation $(l^*, r^*)$ by simulation requires significant computations. Therefore, we will try to estimate the redundancy allocation from \eqref{eq:BDEC_opt_problem_estimate} instead of the simulation for estimating the optimal redundancy allocation.

\begin{figure}[!t]
   \centering
   \includegraphics[width=0.60\textwidth]{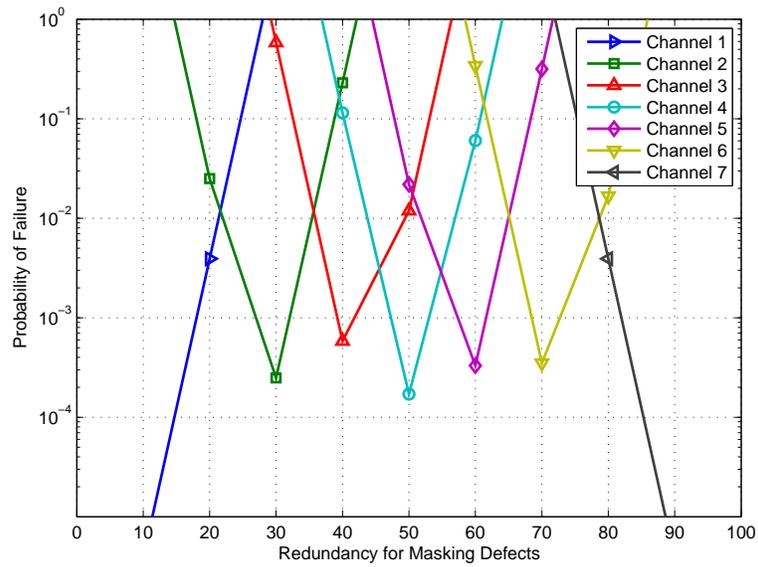}
   \caption{Upper bound on probability of failure for the channels in Table~\ref{tab:channel}.}
   \label{fig:redundancy_allocation_UB}
\end{figure}

First, we can readily obtain the $( \widehat{l}, \widehat{r} )$ that minimizes the objective function of \eqref{eq:BDEC_opt_problem_estimate} for each channel since there only 11 redundancy allocation candidates in Table~\ref{tab:plbc}. The estimate $\widehat{l}$ for all channels can be obtained from Fig.~\ref{fig:redundancy_allocation_UB}. The estimate $\widehat{l}$ for all channels are presented in the third column of Table~\ref{tab:redundancy}. Note that $\widehat{r} = n - k - \widehat{l}$. Table~\ref{tab:redundancy} shows that the estimate $(\widehat{l}, \widehat{r})$ matches the optimal redundancy allocation $(l^*, r^*)$ very well.

Next, $(\widetilde{l}, \widetilde{r})$ can be calculated by \eqref{eq:BDEC_opt_sol_con1}$\sim$\eqref{eq:BDEC_opt_sol_2} assuming that $l$ and $r$ are real values. The solution $\widetilde{l}$ are presented in the last column of Table~\ref{tab:redundancy}. Table~\ref{tab:redundancy} shows that the optimal $l^*$ is the nearest one from $\widetilde{l}$ considering the possible redundancy allocation candidates in Table~\ref{tab:plbc}

\begin{table}[t]
\renewcommand{\arraystretch}{1.3}
\caption{Optimal Redundancy Allocation $l^*$ and its Estimate $\widehat{l}$ and $\widetilde{l}$}
\label{tab:redundancy}
\centering
{\small
\begin{tabular}{|c|c|c|c|}
\hline
Channel & {$l^*$} & {$\widehat{l}$} & {$\widetilde{l}$}    \\ \hline \hline
1 & 0  & 0  & 0  \\ \hline
2 & 30 & 30 & 28.4  \\ \hline
3 & 40 & 40 & 42.8 \\ \hline
4 & 50 & 50 & 50.5  \\ \hline
5 & 60 & 60 & 58.1  \\ \hline
6 & 70 & 70 & 72.2  \\ \hline
7 & 100& 100& 100 \\ \hline
\end{tabular}}
\end{table}

\section{Conclusions}\label{sec:conclusion}

The duality of erasures and defects was revealed. The erasures are corrected by the decoder and the defects are masked by the encoder. The duality holds for channel capacities, capacity achieving schemes, minimum distances, upper bounds on probabilities of failure, and probabilities of failure. By using the upper bounds on the probability of failures, it was proved that the capacities of the BEC and the BDC can be achieved by solving overdetermined linear equations and underdetermined linear equations, respectively.

Also, the BDEC was introduced, which has both erasures and defects. The capacity of the BDEC can be achieved by the coding scheme that combines the coding schemes of the BEC and the BDC.

In addition, we investigated the redundancy allocation for the BDEC. The optimal redundancy allocation was obtained by simulations. In order to reduce the computation complexity, we proposed two methods to estimate the optimal redundancy allocation based on the upper bound on failure probability. The numerical results showed that the estimates of redundancy allocation match the optimal redundancy allocation well.

\appendix[Derivation of $(\widetilde{l}, \widetilde{r})$]

Assume that $l$ and $r$ are real values. Since the objective function is convex and other constraints are linear, the optimization problem of \eqref{eq:BDEC_opt_problem_estimate} is convex. The Lagrangian $L$ is given by
\begin{equation}
\begin{aligned}
L\left( l, r, \lambda_1, \lambda_2, \lambda_3, \lambda_4, \nu \right) & = 2^{-l} \left(1 + \beta \right)^n + 2^{-r} \left\{ 1 + \alpha \left(1 - \beta \right) \right\}^n \\
& + \lambda_1 (-l) + \lambda_2 (-r) + \lambda_3 \left\{l - (n - k)\right\} + \lambda_4 \left\{r - (n - k)\right\} \\
& + \nu \left\{ l + r - (n - k) \right\}
\end{aligned}
\end{equation}
where $\lambda_i$ for $i = 1, 2, 3, 4$ are the Lagrange multipliers associated with the inequality constraints and $\nu$ is the Lagrange multiplier with the equality constraint \cite{Boyd2004convex}.

The KKT conditions are as follows.
\begin{align}
\nabla L &= 0 \label{eq:appendix_KKT_gradient}\\
-l & \le 0 \\
-r & \le 0 \\
l - (n - k) & \le 0 \\
r - (n - k) & \le 0 \\
l + r - (n - k ) & = 0 \label{eq:appendix_KKT_equality} \\
\lambda_i & \ge 0, \quad i = 1,\ldots, 4 \\
\lambda_1 l & = 0 \\
\lambda_2 r & = 0 \\
\lambda_3 \left\{ l - (n - k) \right\} & = 0 \\
\lambda_4 \left\{ r - (n - k) \right\} & = 0
\end{align}
where \eqref{eq:appendix_KKT_gradient} is given by
\begin{equation}\label{eq:appendix_KKT_gradient_expand}
\nabla L =
\begin{bmatrix}
- \ln{2} \cdot 2^{-l} \left( 1 + \beta \right)^n  \\
- \ln{2} \cdot 2^{-r} \left\{ 1 + \alpha \left( 1 - \beta \right) \right\}^n
\end{bmatrix}
+
\begin{bmatrix}
-\lambda_1 + \lambda_3 + \nu \\
-\lambda_2 + \lambda_4 + \nu
\end{bmatrix}
= 0.
\end{equation}

We will consider the following three conditions:
\begin{itemize}
  \item $l=0, r=n-k$
  \item $l=n-k, r=0$
  \item $0 < l < n-k, 0 < r < n-k$
\end{itemize}

1) $l=0, r=n-k$

Due to \emph{complementary slackness}, it is clear that $\lambda_2 = \lambda_3 = 0$. Thus, \eqref{eq:appendix_KKT_gradient_expand} will be as follows.
\begin{align}
- \ln{2} \cdot 2^{-l} \left( 1 + \beta \right)^n - \lambda_1 +  \nu & = 0 \\
- \ln{2} \cdot 2^{-r} \left\{ 1 + \alpha \left( 1 - \beta \right) \right\}^n + \lambda_4 + \nu  & = 0
\end{align}
Since $\lambda_1 + \lambda_4 \ge 0$,
\begin{equation} \label{eq:appendix_opt_sol_con1}
2^{-l} \left(1 + \beta \right)^n \le 2^{-r} \left\{ 1 + \alpha \left(1 - \beta \right) \right\}^n.
\end{equation}
which results in \eqref{eq:BDEC_opt_sol_con1}. It reveals that we have to allot all redundancy for correcting erasures if \eqref{eq:appendix_opt_sol_con1} is true.

2) $l=n-k, r=n-k$

Due to complementary slackness, it is clear that $\lambda_1 = \lambda_4 = 0$. Thus, \eqref{eq:appendix_KKT_gradient_expand} will be as follows.
\begin{align}
- \ln{2} \cdot 2^{-l} \left( 1 + \beta \right)^n + \lambda_3 +  \nu & = 0 \\
- \ln{2} \cdot 2^{-r} \left\{ 1 + \alpha \left( 1 - \beta \right) \right\}^n - \lambda_2 + \nu  & = 0
\end{align}
Since $\lambda_2 + \lambda_3 \ge 0$,
\begin{equation} \label{eq:appendix_opt_sol_con2}
2^{-l} \left(1 + \beta \right)^n \ge 2^{-r} \left\{ 1 + \alpha \left(1 - \beta \right) \right\}^n.
\end{equation}
which results in \eqref{eq:BDEC_opt_sol_con2}. It reveals that we have to allot all redundancy for masking defects if \eqref{eq:appendix_opt_sol_con2} is true.

3) $0 < l < n-k, 0 < r < n-k$

Due to complementary slackness, it is clear that $\lambda_1 = \lambda_2 = \lambda_3 = \lambda_4 = 0$. Thus, \eqref{eq:appendix_KKT_gradient_expand} will be as follows.
\begin{align}
- \ln{2} \cdot 2^{-l} \left( 1 + \beta \right)^n + \nu & = 0 \label{eq:appendix_opt_sol_con3_1} \\
- \ln{2} \cdot 2^{-r} \left\{ 1 + \alpha \left( 1 - \beta \right) \right\}^n + \nu  & = 0 \label{eq:appendix_opt_sol_con3_2}
\end{align}
By \eqref{eq:appendix_opt_sol_con3_1} and \eqref{eq:appendix_opt_sol_con3_2},
\begin{equation} \label{eq:appendix_opt_sol_con3_3}
2^{-l} \left( 1 + \beta \right)^n = 2^{-r} \left\{ 1 + \alpha \left( 1 - \beta \right) \right\}^n.
\end{equation}
By \eqref{eq:appendix_KKT_equality} and \eqref{eq:appendix_opt_sol_con3_3}, \eqref{eq:BDEC_opt_sol_1} and \eqref{eq:BDEC_opt_sol_2} can be obtained.


%

%
%

%
%

\ifCLASSOPTIONcaptionsoff
  \newpage
\fi



\bibliographystyle{IEEEtran}
\bibliography{IEEEabrv,duality_v5}

\newpage

\end{document}